\documentclass[fleqn,usenatbib]{mnras}

\usepackage{newtxtext,newtxmath}

\usepackage[T1]{fontenc}
\usepackage{ae,aecompl}

\usepackage{graphicx}	
\usepackage{amsmath}	
\usepackage{amssymb}	
\usepackage{hyperref}
\usepackage{natbib}
\usepackage[caption=false]{subfig}
\usepackage{ragged2e}
\usepackage{placeins}
\usepackage{bigints} 
\usepackage{physics}
\usepackage{xcolor}
\usepackage{cleveref}
\usepackage[export]{adjustbox}

\def\mean#1{\left< #1 \right>}

\title[Stratified turbulence in the ICM]{Turbulence in stratified atmospheres: implications for the intracluster medium}

\author[Mohapatra, Federrath \& Sharma]{
Rajsekhar Mohapatra,$^{1}$\thanks{E-mail: rajsekhar.mohapatra@anu.edu.au}
Christoph Federrath,$^{1}$
Prateek Sharma$^{2}$
\\
$^{1}$Research School of Astronomy and Astrophysics, Australian National University, Canberra, ACT 2611, Australia\\
$^{2}$Department of Physics, Indian Institute of Science, Bangalore, 560012, India\\
}

\date{Accepted XXX. Received YYY; in original form ZZZ}

\pubyear{2019}

\begin{document}
\label{firstpage}
\pagerange{\pageref{firstpage}--\pageref{lastpage}}
\maketitle

\begin{abstract}
The gas motions in the intracluster medium (ICM) are governed by turbulence. However, since the ICM has a radial profile with the centre being denser than the outskirts, ICM turbulence is stratified.
Stratified turbulence is fundamentally different from Kolmogorov (isotropic, homogeneous) turbulence; 
kinetic energy not only cascades from large to small scales, but it is also converted into buoyancy potential energy.
To understand the density and velocity fluctuations in the ICM, we conduct high-resolution ($1024^2\times 1536$ grid points) hydrodynamical simulations of subsonic turbulence (with rms Mach number $\mathcal{M}\approx 0.25$) and different levels of stratification, quantified by the Richardson number $\mathrm{Ri}$, from $\mathrm{Ri}=0$ (no stratification) to $\mathrm{Ri}=13$ (strong stratification).
 We quantify the density, pressure and velocity fields for varying stratification because observational studies often use surface brightness fluctuations to infer the turbulent gas velocities of the ICM. 
We find that the standard deviation of the logarithmic density fluctuations ($\sigma_s$), where $s=\ln(\rho/\mean{\rho(z)})$, 
increases with $\mathrm{Ri}$. For weakly stratified subsonic turbulence ($\mathrm{Ri}\lesssim10$, $\mathcal{M}<1$), we derive a new \mbox{$\sigma_s$--$\mathcal{M}$--$\mathrm{Ri}$} relation,  $\sigma_s^2=\ln(1+b^2\mathcal{M}^4+0.09\mathcal{M}^2 \mathrm{Ri} H_P/H_S)$, where \mbox{$b=1/3$--1} is the turbulence driving parameter, and $H_P$ and $H_S$ are the pressure and entropy scale heights respectively. 
We further find that the power spectrum of density fluctuations, $P(\rho_k/\mean{\rho})$, increases in magnitude with increasing $\mathrm{Ri}$. Its slope in $k$-space flattens with increasing $\mathrm{Ri}$ before steepening again for $\mathrm{Ri}\gtrsim1$. In contrast to the density spectrum, the velocity power spectrum is invariant to changes in the stratification. Thus, we find that the ratio between density and velocity power spectra strongly depends on $\mathrm{Ri}$, with the total power in density and velocity fluctuations described by our \mbox{$\sigma_s$--$\mathcal{M}$--$\mathrm{Ri}$} relation. Pressure fluctuations, on the other hand, are independent of stratification and only depend on $\mathcal{M}$.

\end{abstract}




\section{Introduction}\label{sec:introduction}

\begin{figure}
	\centering
	\includegraphics[width=0.99\columnwidth]{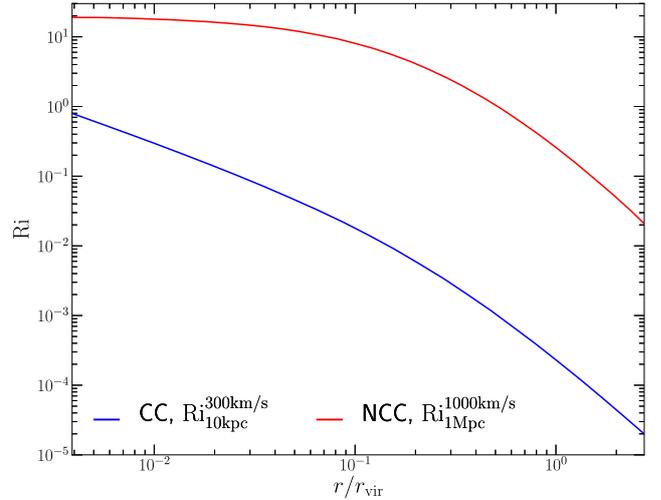}	
	\caption[Ri-cluster plots]{Typical $\mathrm{Ri}$ profiles for cool core (CC) and non-cool core (NCC) clusters. We calculate $\mathrm{Ri}$ using the driving length scale $L_{\mathrm{driv}}$ and the turbulent velocity on scale $L_{\mathrm{driv}}$, $v_{L_{\mathrm{driv}}}$, indicated in the subscripts and superscripts in the legends, respectively. For CC clusters, we consider driving by AGN jets with $L_{\mathrm{driv}}\approx 10$ $\mathrm{kpc}$ and $v_{L_{\mathrm{driv}}}\approx300$ $\mathrm{km/s}$ \citep{hitomi2016}. For NCC clusters, we consider driving by mergers with $L_{\mathrm{driv}}\approx 1$ $\mathrm{Mpc}$ and $v_{L_{\mathrm{driv}}}\approx1000$ $\mathrm{km/s}$. Here we use the Brunt-Vaisala frequency $N$ from figure~1 in \cite{Shi2019}. Note that $\mathrm{Ri}=N^2L_{\mathrm{driv}}^2/v_{L_{\mathrm{driv}}}^2$ is sensitive to the choice of $L_{\mathrm{driv}}$ and $v_{L_{\mathrm{driv}}}$. Moderately stratified turbulence with $\mathrm{Ri}\lesssim10$ governs both cool and non-cool core clusters.}
	\label{fig:cluster-Ri-profile}
\end{figure}

An interplay between turbulence and gravity is observed in several terrestrial and astrophysical fluid systems, starting from the earth's atmosphere and oceans to hot ionised plasma in clusters of galaxies \citep{Stein1967,Low1988,Fernando1996,Parmentier2013}. All these phenomena are governed by the physics of stratified turbulence. In our study we focus on stratified turbulence in the regimes relevant to the intracluster medium (ICM), which is moderately stratified in the gas density, i.e., the ratio of buoyancy and the turbulent shear forces on the driving scale - defined as the Richardson number ($\mathrm{Ri}$) is $\lesssim 10$. Based on their entropy profiles, clusters can be broadly divided into cool cores (CCs) and non-cool cores (NCCs). CC clusters are relaxed with a high (low) central gas density (entropy), and NCCs have large velocity dispersions and are typically undergoing mergers. In \Cref{fig:cluster-Ri-profile} we show the typical $\mathrm{Ri}$ profile for these two types of clusters.



Recent observations \citep{zhuravleva2014turbulent,hitomi2016,zhuravleva2018} have tried to measure turbulent velocities in cluster cool cores, which is important to infer details about gas motions and thermodynamics. \citeauthor{hitomi2016} measured the gas velocities directly by looking at Doppler line-broadening of Fe\texttt{xxv} and Fe\texttt{xxvi} lines, but such direct measurements will only be available in a few years.\footnote{\url{https://global.jaxa.jp/projects/sas/xrism/}}  We can infer the turbulent velocities indirectly if we can relate them to density fluctuations, which are easier to measure from X-ray surface brightness maps (e.g., \citealt{zhuravleva2014turbulent}). Other than cool cores, on larger scales such as cluster outskirts, turbulent pressure is a key component of the non-thermal pressure support. Estimating turbulent velocities is again important in calculating the hydrostatic mass bias of clusters \citep{schuecker2004,George2009,Bautz2009,Cavaliere2011,Nelson2014ApJ}. On these scales, pressure fluctuations obtained from Sunyaev-Zeldovich effect (SZ) observations can also be used to estimate gas velocities \citep{Zeldovich1969,khatri2016,Mroczkowski2019}. Thus, a detailed study of density, pressure and velocity fluctuations in a stratified medium is important to obtain reliable scaling relations between different observables and velocities (see \citealt{Simionescu2019SSRv} for a review on ICM gas velocities).

Many current turbulence studies of the ICM typically ignore the effects of gravity (and hence stratification), and model it using homogeneous isotropic turbulence \citep{Brunetti2007,banerjee2014turbulence,Mohapatra2019,Grete2020}. However, the gas distribution in the ICM is neither homogeneous nor isotropic. Instead, the density distribution is stratified and is in rough hydrostatic equilibrium with the gravitational profile.  Stratified turbulence itself is fundamentally different from homogeneous isotropic turbulence, in that it opens up a new channel of energy exchange. While in homogeneous isotropic turbulence, kinetic energy cascades down through eddies of decreasing sizes and is ultimately dissipated on the viscous length scale \citep{Frisch1995}, in stratified turbulence energy can also be channeled back and forth between kinetic energy and gravitational potential energy \citep{Bolgiano1962,verma2018}. Due to these fundamental differences $\epsilon_{\ell}$, the rate of transfer of kinetic energy on length scale $\ell$ can vary as a function of $\ell$. This can change the interpretation of important statistical tools used to analyse turbulence, such as velocity and density distributions, power spectra of velocity and 
density and their correlation functions. 

In most current theoretical and computational studies of stratified turbulence, driving is generally perpendicular to the direction of gravity.  \citep{Carnevale2001,Lindborg2006,Brethouwer2008,Herring2013,Kumar2014}. They are designed to model stratified turbulence in planetary atmosphere and oceans. However turbulence in the cluster cores is driven more isotropically by active galactic nuclei (AGN) jets and gravitational mergers \citep{Balbus1990,Churazov2002MNRAS,Churazov2003,Omma2004MNRAS,Nelson2012ApJ}. Also, to obtain scaling relations between density and velocity fluctuations we need to scan a larger parameter space of $\mathrm{Ri}$ than what has been done in current studies. We therefore drive turbulence isotropically in a local idealised model of a cluster atmosphere. We present scaling relations between density, pressure and velocity fluctuations to understand stratified turbulence through our simulations.

Recent ICM turbulence studies such as \citet{Gaspari2012,zhuravleva2014relation,Zhang2018,Valdarnini2019,Shi2019} include gravitational stratification in their models. However, these 
simulations mostly cannot resolve the Ozmidov length  $l_O$ (the length scale where turbulent shear and stratification terms are equal in magnitude) within the inertial range of turbulence. Even in high resolution simulations where $l_O$ is resolved, there is a lack of a detailed study of how turbulence varies with the changing strength of stratification in clusters as parameterised by $\mathrm{Ri}$.

In this study, we analyse turbulence by varying the stratification strength in the parameter space relevant to the ICM (i.e. $0.001\lesssim\mathrm{Ri}\lesssim10$, see \Cref{tab:sim_params}) and compare some key turbulence statistics (such as probability distribution functions (PDFs) of velocity, density, power spectra of density and velocity). We resolve $l_O$ in the inertial range of our simulations.

The paper is organised as follows: in \cref{sec:Methods} we describe our setup and methods, we present our results and their interpretations in \cref{sec:Results}, we compare our results with the literature and discuss the caveats of our work in \cref{sec:caveats-future}, and we conclude in \cref{sec:Conclusion}. 

\section{Methods}\label{sec:Methods}
\subsection{Model equations}\label{subsec:ModEq}
We model the ICM using hydrodynamic equations with gravity and turbulent forcing as additional source terms in the momentum and energy equations.
We solve the following equations:
\begin{subequations}
	\begin{align}
	\label{eq:Euler1}
	&\frac{\partial\rho}{\partial t}+\nabla\cdot (\rho \mathbf{v})=0,\\
	\label{eq:Euler2}
	&\frac{\partial(\rho\mathbf{v})}{\partial t}+\nabla\cdot (\rho \mathbf{v}\otimes \mathbf{v})+\nabla P=\rho\mathbf{F}+\rho \mathbf{g},\\
	\label{eq:Euler3}
	&\frac{\partial E}{\partial t}+\nabla\cdot ((E+P)\mathbf{v})=\rho\mathbf{F}\cdot\mathbf{v}+\rho (\mathbf{v}\cdot\nabla)\Phi,
	\end{align}
\end{subequations}
where $\rho$ is the gas mass density, $\mathbf{v}$ is the velocity, $P=\rho k_B T/(\mu m_p)$ is the pressure (we use the ideal gas equation of state), $\mathbf{F}$ is the turbulent acceleration that we apply, $\Phi$ is the gravitational potential, $\mathbf{g}=-\nabla\Phi$ is the acceleration due to gravity, $E=\rho v^2/2 + P/(\gamma-1)$ 
is the sum of kinetic and internal energy densities, $\mu$ is the mean particle weight, $m_p$ is the proton mass, $k_B$ is the Boltzmann constant, $T$ is the temperature, and $\gamma=5/3$ is the adiabatic index. We assume that the gas is fully ionised with a third solar metallicity (which is the case for typical ICM conditions) and this gives us $\mu\approx0.61$. 
\subsection{Setup}\label{subsec:Setup}
We choose $-\hat{\mathbf{z}}$ to be the direction of the gravitational field, and pressure and density to have scale heights $H_P$ and $H_{\rho}$, respectively. Thus, the initial pressure and density profiles are given by
 \begin{subequations}
 	\begin{align}
 	\label{eq:Presinitial}
 	&P(t=0)=P_0\exp(-\frac{z}{H_P}) \text{ and}\\
 	\label{eq:Densinitial}
 	&\rho(t=0)=\rho_0\exp(-\frac{z}{H_{\rho}})\text{, respectively.} 	
 	\end{align}
 \end{subequations}
Since we start with the gas in hydrostatic equilibrium, the initial density and pressure are related by 
 \begin{subequations}
\begin{equation}\label{eq:grav}
 	\frac{\mathrm{d}P}{\mathrm{dz}}=-\rho g. 
\end{equation}
 Hence $g$ is set as:
	\begin{align}
		\label{eq:grav1}
		g&=\frac{P_0}{\rho_0 H_P}\exp(-z\left[\frac{1}{H_P}-\frac{1}{H_{\rho}}\right]).
	\end{align} 
 \end{subequations}
This equilibrium is convectively stable if $\mathrm{d}\ln S/\mathrm{dz}>0$, where 
\begin{equation}
S=\frac{P}{\rho^{\gamma}}\text{ is the pseudo-entropy.}
\end{equation} This gives us the condition for the entropy scale height $H_S$ ($\equiv 1/[\mathrm{d} \ln S/\mathrm{d}z]$)
\begin{equation}\label{eq:H_S}
	\frac{1}{H_S}=\frac{\gamma}{H_{\rho}}-\frac{1}{H_P}>0.
\end{equation}
This condition is satisfied for all our simulations, which locally mimic the stably stratified ICM.

\begin{table*}
	\centering
	\caption{Simulation parameters for different runs}
	\label{tab:sim_params}
	\resizebox{\textwidth}{!}{
		\begin{tabular}{lcccc|ccc} 
			\hline
			Label & $\mathrm{Ri}$ & $H_{\rho}$ & $H_P$ & Resolution  &  $\mathcal{M}$ & $\sigma_s$ & $\alpha$\\
			(1) & (2) & (3) & (4) & (5) & (6) & (7) & (8)\\
			\hline
			Ri0.05LowRes & $0.05\pm0.02$ & $3.0$ &  $6.0$ & $512^2\times768$ & $0.254\pm0.008$ & $0.035\pm0.005$ & $-1.71\pm0.06$\\
			\hline
			Ri0 & $0$ & $\infty$ &  $\infty$ & $1024^2\times1536$ & $0.26\pm0.01$ & $0.021 \pm 0.001$  & $-2.06 \pm 0.01$\\
			Ri0.003 & $0.003\pm0.001$ & $15.0$ &  $30.0$ & $1024^2\times1536$ & $0.26\pm0.01$ & $0.022 \pm 0.001$ & $-1.99 \pm 0.02$\\
			Ri0.01 & $0.012\pm0.003$ & $6.0$ &  $12.0$ & $1024^2\times1536$ & $0.256\pm0.002$ & $0.026 \pm 0.002$ & $-1.96 \pm 0.02$ \\
			Ri0.05 &  $0.05\pm0.02$ & $3.0$ &  $6.0$ & $1024^2\times1536$ & $0.258\pm0.007$ & $0.037 \pm 0.003$ & $-1.86 \pm 0.02$\\
			Ri0.2 &  $0.24\pm0.06$ & $1.5$ &  $3.0$ & $1024^2\times1536$ & $0.25\pm0.01$ & $0.06\pm0.01$  & $-1.79\pm0.03$\\
			Ri0.5 & $0.5\pm0.1$  & $1.0$ &  $2.0$ & $1024^2\times1536$ & $0.25\pm0.02$ & $0.08 \pm 0.01$ & $-1.77 \pm 0.04$\\
			Ri0.6NV & $0.6\pm0.2$ & $1.0$ &  $2.0$ & $1024^2\times1536$ & $0.25\pm0.01$ & $0.07\pm0.01$ & $-1.77 \pm 0.03$\\
			Ri1 & $1.0\pm0.3$ & $0.75$ &  $1.5$ & $1024^2\times1536$ & $0.25\pm 0.01$ & $0.10\pm0.02$  & $-1.77\pm0.03$\\
			Ri3 & $3.0\pm1.0$ & $0.5$ &  $1.0$ & $1024^2\times1536$ & $0.25\pm 0.01$ & $0.14\pm0.01$  & $-1.83\pm	0.03$\\
			Ri13 & $13^{-6}_{+11}$ & $0.25$ &  $0.5$ & $1024^2\times1536$ & $0.26\pm 0.02$ & $0.20\pm0.02$ & $-1.91\pm0.01$\\
			Ri10NV & $9^{-4}_{+7}$  & $0.25$ &  $0.5$& $1024^2\times1536$ & $0.30\pm0.01$ & $ 0.22\pm0.02$ & $-1.88\pm0.02$\\
			\hline
			Ri0.05HighRes & $0.05\pm0.02$ & $3.0$ &  $6.0$ &  $2048^2\times3072$ & $0.257\pm0.007$  & $0.037\pm0.003$ & $-1.86\pm0.01$ \\			
		\hline
			
	\end{tabular}}
	\justifying \\ \begin{footnotesize} Notes: Column 1 shows the simulation name. The number following `$\mathrm{Ri}$' is the average Richardson number in the simulation. The default resolution of all runs is $1024^2 \times 1536$, `LowRes' refers to $512^2 \times 768$ and `HighRes' to $2048^2 \times 3072$. `NV' stands for turbulent forcing only in the directions perpendicular to gravity. In columns~2, 3 and 4, we list $\mathrm{Ri}$, $H_{\rho}$, and $H_P$, which are simulation parameters defined in \cref{eq:Ri,eq:Presinitial,eq:Densinitial}. Column~5 lists the grid resolution. In column 5, $\mathcal{M}$ refers to the rms Mach number (see equation~\ref{eq:mach}). Column~7 shows $\sigma_s$, the standard deviation of $s$ (defined in equation~\ref{eq:s_log_rho}). In column 8, $\alpha$ refers to the spectral slope of $P(\bar{\rho}_k)$, the power spectrum of normalised density fluctuations. We have defined the normalisation of density fluctuations in \cref{eq:bar_rho}. All quantities ($\mathcal{M}$, $\alpha$, $\sigma_s$) were averaged over 5 turbulent turnover times, for $3 \leq t/t_\mathrm{turb} \leq 8$.\end{footnotesize} 
	
\end{table*}

\subsection{Calculating the Richardson number ($\mathrm{Ri}$)}\label{subsec:CalcRi}
 Consider a parcel of gas with density $\rho^\prime$ and pressure $P^\prime$ at $z=z$, moving with $v_z>0$. As the parcel rises subsonically to $z+\delta z$, it is in pressure equilibrium with its surroundings, but does not exchange energy with it. Thus the gas parcel is adiabatic. The Lagrangian changes in the pressure and density of the parcel, respectively, are
 \begin{subequations}
 	\begin{align}
 		\Delta P^\prime&=P^\prime(z+\delta z)-P^\prime(z)=\frac{\mathrm{d}P^\prime}{\mathrm{d}z}\delta z,\\
 		\Delta\rho^\prime&=\left(\frac{\partial \rho^\prime}{\partial P^\prime}\right)_S\Delta P^\prime.
 	\end{align}
 	Since the ambient density $\rho(z)$ itself changes as a function of $z$, the Eulerian overdensity at $z+\delta z$ is given by:
 	\begin{align}
 		\delta \rho =\frac{\mathrm{d}P}{\mathrm{d}z}\left[\left(\frac{\partial \rho^\prime}{\partial P^\prime}\right)_S - \frac{\mathrm{d}\rho}{\mathrm{d}P}\right]\delta z.\label{eq:dens_buoy}
 	\end{align}
The quantity inside the square brackets should be positive for a stable stratification. The buoyancy acceleration acting on the displaced parcel is $\delta \rho^\prime g/\rho$ and the equation of motion for the gas parcel is
\begin{align}\label{eq:EoM}
	\frac{\mathrm{d}^2\delta z}{\mathrm{d}t^2}&=  -\underbrace{\frac{g}{\gamma} \frac{d}{dz} \ln \left (\frac{P}{\rho^\gamma} \right)}_{N^2}\delta z = -N^2 \delta z,
\end{align}
 \end{subequations}
where $N^2$ is the square of the Brunt-Vaisala (BV) frequency. 
The condition for stable stratification is $N^2>0$, or
	\begin{align}
		&\left(\frac{\gamma}{H_{\rho}}-\frac{1}{H_P}\right)>0, \text{ or }H_{\rho}/H_P<\gamma,
	\end{align}as mentioned earlier.

The scale-dependent turbulent Richardson number $\mathrm{Ri_{\ell}}$ at a length scale $\ell$ is defined as
\begin{equation}
\label{eq:Ri}
	\mathrm{Ri}_{\ell}=\frac{N^2}{(v_{\ell}/\ell)^2},
\end{equation}
where $v_{\ell}$ is the velocity on length scale $\ell$.  On large scales, stratification is expected to be dominant over turbulence so we can have $\mathrm{Ri}_{\ell}>1$, while on small enough length scales, turbulence dominates to give $\mathrm{Ri}_{\ell}<1$. The length scale on which $\mathrm{Ri}_{\ell}=1$ is called the Ozmidov length scale $l_{\mathrm{O}}$ ($\mathrm{Ri}_{l_{\mathrm{O}}}=1$).

\Cref{tab:sim_params} shows the different choices of $H_{\rho}$ and $H_P$ for different simulations and the Richardson number $\mathrm{Ri}_{L_{\text{driv}}}$ on the driving scale $L_{\text{driv}}$, which we shall refer to as $\mathrm{Ri}$ from now on. An equivalent dimensionless number, the Froude number ($\mathrm{Fr} = \mathrm{Ri}^{-1/2}$) is sometimes used to describe turbulence in stratified fluids.
\subsection{Normalisation}\label{subsec:normalisation}
In order to compare density, pressure and velocity fluctuations in stably stratified turbulence, we first normalise these fluctuations to construct dimensionless variables $\bar{\rho}$, $\bar{P}$ and $\mathcal{M}$ and $s$. These variables are given by
\begin{subequations}
\begin{align}
&\bar{\rho}=\frac{\rho}{\mean{\rho(z)}}\text{,}\label{eq:bar_rho}\\
&\bar{P}=\frac{P}{\mean{P(z)}}\text{,}\label{eq:bar_P}\\
&\mathcal{M}=\mean{\frac{v}{c_s}}_{\mathrm{rms}}\text{ and}\label{eq:mach}\\
&s=\ln(\bar{\rho})\text{,}\label{eq:s_log_rho}
\end{align}
where $\mean{\rho(z)}$ and $\mean{P(z)}$ are the average density and pressure at $z=z$ slice, respectively, $v$ is the amplitude of velocity and $c_s$ is the local speed of sound. 
\end{subequations}
\subsection{Calculating the potential energy ($E_{u_b}$)}\label{subsec:CalcPE}
\Cref{eq:EoM} shows that a gas parcel experiences a restoring force when displaced in the vertical direction.
In the absence of continuous turbulence driving, this parcel would oscillate about its mean position like a harmonic oscillator with a natural frequency $N$ \citep[Chapter 4,][]{Lighthill1978}. 
The potential energy (per unit mass) is thus given by 
\begin{align}
	N^2 \delta z^2/2 = g^2 (\delta \rho/\rho)^2/2 N^2,
\end{align} 
using \cref{eq:dens_buoy,eq:EoM}. We define a quantity $u_b=g\delta\bar{\rho}/N$ with dimensions of velocity. The potential energy per unit mass is given by
\begin{equation}\label{eq:epot}
	E_{u_b}=u_b^2/2.
\end{equation}
On substituting the expressions for $g$ and $N$ from \cref{eq:grav1,eq:EoM} for the two profiles, we obtain
\begin{equation}
E_{u_b}=\frac{P}{2\rho}\frac{\delta\bar{\rho}^2}{\frac{H_P}{H_{\rho}}-\frac{1}{\gamma}}.
\end{equation}
As usual, the fluctuating kinetic energy per unit mass is $E_u = \delta v^2/2$, where $\delta v$ is the magnitude of the fluctuating velocity at a given height.
\subsection{Numerical methods}
\label{subsec:Num_Methods}
We use a modified version of the FLASH code \citep{Fryxell2000,Dubey2008}, version 4, for our simulations. We evolve the Euler equations in FLASH, with additional gravity and forcing terms in the momentum and energy equations (\ref{eq:Euler1} to \ref{eq:Euler3}). The pressure, density and temperature of the gas are related by the ideal gas equation of state. We use the hydrodynamic version of the HLL3R Riemann solver \citep{Bouchut2007,Bouchut2010} to solve the Euler equations. This solver has been tested for efficiency, robustness and accuracy in \cite{Waagan2011}. We use a uniformly spaced 3D grid, with periodic boundary conditions along the $x$ and $y$ directions. In the $z$ direction, we use reflective boundary conditions. We work with dimensionless units, such that density ($\rho_0$) and initial speed of sound ($c_{s,0}$) at $z=0$ are unity (refer to equations \ref{eq:Presinitial} and \ref{eq:Densinitial} for definitions).
We set the box size along $x$ and $y$ to be $L_x=L_y=1$. In order to minimise any anomalous effects of the reflective boundary conditions, we use a larger box size along the $z$ direction, $L_z=1.5$. However, we analyse the simulations only inside the cube of size $L=1$ centred at $(0,0,0)$. Thus, the boundaries of our analysis box are at $(\pm0.5,\pm0.5,\pm0.5)$. Our simulations have a maximum grid resolution of $2048^2\times3072$ (with most runs being $1024^2 \times 1536$), the larger number of cells being in the $z$ direction, so that the individual cells are cubical. We have checked our results for convergence over different resolutions ($512^2 \times 768$, $1024^2 \times 1536$ and $2048^2 \times 3072$)  in \Cref{fig:PDF-convergence-120,fig:convergence-120}. We have confirmed that any effects of our choice of boundary conditions and box size along the $z$ direction are minimal, by testing other boundary conditions and different box sizes along the $z$ direction.

\subsection{Turbulent forcing}
\label{subsec:Turb_forcing}

We follow a spectral forcing method using the stochastic Ornstein-Uhlenbeck (OU) process to model the turbulent acceleration $\mathbf{F}$ with a finite autocorrelation time scale $t_{\mathrm{turb}}$ \citep{eswaran1988examination,schmidt2006numerical,federrath2010}. The acceleration $\mathbf{F}$ only contains large-scale modes, $1\leq|\mathbf{k}|L/2\pi\leq3$. The power injected is a parabolic function of  $|\mathbf{k}|$ and peaks at the $k_{\text{inj}}=2$ mode in Fourier space, i.e.~on half of the box size (for simplicity, we have dropped the wavenumber unit $2\pi/L$ in the rest of the paper). Turbulence on scales $k\geq3$ develops self-consistently, and is not directly driven. We set the autocorrelation time-scale on the driving scale, $t_{\mathrm{turb}}=(L/2)/\sigma_v$, where $\sigma_v$ is the standard deviation of velocity on the driving scale, $L/2$. Our driving is solenoidal, i.e., the acceleration field has zero divergence and non-zero curl. Compressive forcing can give rise to larger density perturbations \citep{Federrath2008}, which we do not include in this study.  We use the same forcing field for all of our simulations. For further details of the forcing method, refer to section 2.1 of \cite{federrath2010}. 

To minimise any anomalous effects of the reflective boundary conditions in the $z$ direction, we apply a window function on the turbulence acceleration field. This window function slowly decays the external acceleration amplitude to zero close to the reflective boundaries in the $z$ direction (for all three components of acceleration). The window function is given by
\begin{align}
w(z)&=1 \text{ for }\abs{z}<0.625,\nonumber\\
       &=\exp(-\abs{\abs{z}-0.625}/0.125) \text{ for }\abs{z}>0.625.
\end{align}
Note that $w(z)=1$ inside the analysis box. Thus, the driving amplitude is reduced exponentially, only very close to the computational domain boundaries in $z$.

\subsection{Relevance to the ICM}
\label{subsec:scalingtoICM}
Although we work with dimensionless units, our results can easily be scaled to model the ICM. If we take $L=40\, \mathrm{ kpc}$, Then the driving scale ranges between $10-20\ \mathrm{ kpc}$. If the sound speed $c_s\approx 500\ \mathrm{ km/s}$, then with $\mathcal{M}\approx 0.25$, $v_{\mathrm{turb}}\approx 125\ \mathrm{ km/s}$.  
The gas temperature profile is given by
\begin{align}
T&=T_0\exp(-z\left[\frac{1}{H_P}-\frac{1}{H_{\rho}}\right]).
\end{align}
where $T_0\approx 1 \mathrm{keV}$. The scale height of density $H_\rho$ varies between $10-600 \mathrm{~kpc}$ and $H_P =2H_\rho$ for various runs.
For modelling galaxy clusters, we choose $H_P>H_{\rho}$ such that $T$ increases with increasing $z$. Although the temperature decreases with radius in NCCs, our results are applicable there
too because the buoyant response depends on the entropy gradient rather than the temperature gradient. In the presence of thermal conduction (both isotropic and anisotropic) the physics of stratified turbulence is modified, and temperature fluctuations 
and gradients become important (\citealt{McCourt2011,gaspari2014}). This investigation is beyond the scope of the present paper.

\section{Results and discussion}\label{sec:Results}
Here we describe the results of our simulations and discuss their possible implications in the context of clusters, and stratified turbulence in general. \Cref{tab:sim_params} lists the different simulation parameters for our runs. We use an identical acceleration field to drive turbulence in all our simulations. Thus, all our simulations have approximately the same subsonic $\mathcal{M}\approx0.25$ in steady state. We run the simulations for a total of $8$ eddy turnover times ($t_{\mathrm{eddy}}=t_{\mathrm{turb}}$) on the driving scale. We have chosen a standard set of runs - Ri0, Ri0.003, Ri0.05, Ri0.2, Ri1.0 and Ri13 ($\mathrm{Ri}$ denoted by the number in the label) for our analysis.

\begin{figure}
		\centering
	\includegraphics[width=0.99\columnwidth]{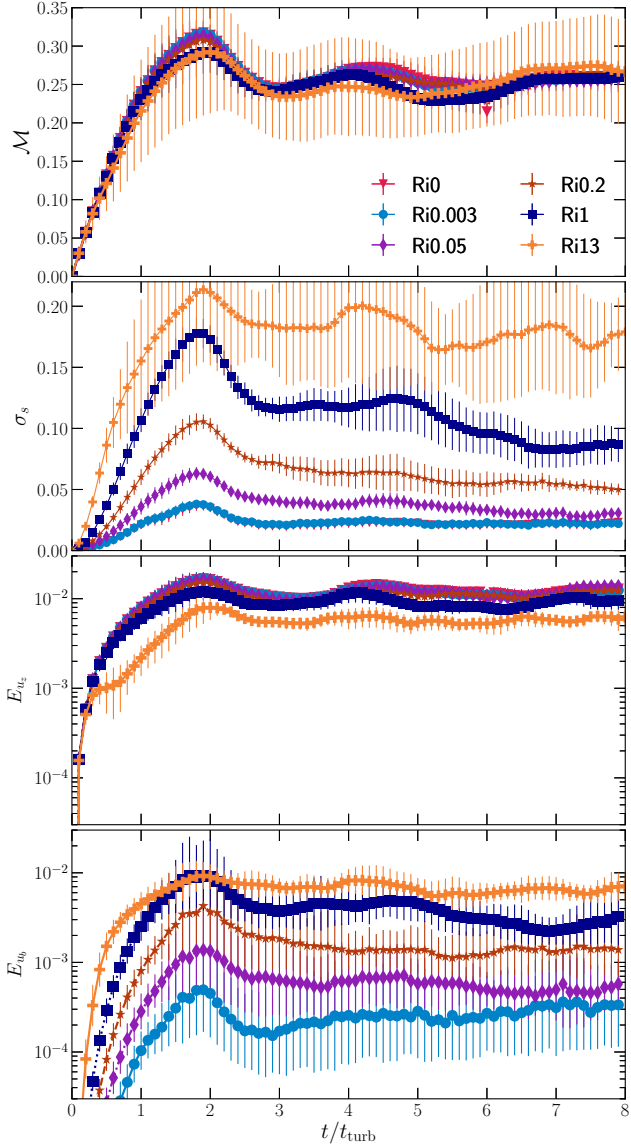}	
	\caption[time-evol plots]{Time evolution of (first panel $\mathcal{M}$) rms Mach number, (second panel $\sigma_s$) logarithmic density fluctuations, (third panel $E_{u_z}$) specific kinetic energy in $z$ direction  and (fourth panel $E_{u_b}$) specific potential energy for the standard set of runs (Ri0, Ri0.003, Ri0.05, Ri0.2, Ri1.0 and Ri13). The errorbars indicate the standard deviation of these quantities after averaging them in constant $z$ slices. In steady state, $\mathcal{M}\approx0.25$ is similar for all runs, $\sigma_s$ and $E_{u_b}$ increase, while $E_{u_z}$ decreases (significantly for $\mathrm{Ri}\gtrsim 1$) with increasing stratification.}
	\label{fig:time-evolution}
\end{figure}

 \Cref{fig:time-evolution} shows the time evolution of $\mathcal{M}$, $\sigma_s$, $E_{u_b}$ and $E_{u_z}$ ($\sigma_s$ in standard deviation in $s$ over the entire analysis cube). All these quantities reach a steady state at around $t=3\ t_{\mathrm{turb}}$. Thus, we start analysing turbulence at this time and we have a total duration of $5\ t_{\mathrm{turb}}$ for statistical averaging (from $3\ t_{\mathrm{turb}}$ to $8\ t_{\mathrm{turb}}$). The gas only heats up slightly due to viscous dissipation of energy, increasing the sound speed with time. However, this change is not significant ($\lesssim10\%$) and $\mathcal{M}$ roughly remains constant for the entire duration of the analysis and statistical averaging, as shown in the top panel of \Cref{fig:time-evolution}.

\subsection{Evolution of density perturbations, and potential and kinetic energies} \label{subsec:dens_amplitude}

X-ray surface brightness fluctuations are used to calculate density perturbations in the ICM, and these are further used to constrain turbulent gas velocities \citep{zhuravleva2014relation}. 
Turbulence alone can produce density fluctuations (e.g., \citealt{Passot1998PhRvE,Federrath2008}). However, with stratification, turbulence gives rise to a new kind of density perturbations. When a parcel of gas moves to higher (lower) $z$, it has higher (lower) density compared to the local profile and appears as positive (negative) density perturbation at the parcel's new location. For unstratified subsonic turbulence, the density perturbations are much smaller and increase roughly quadratically with the Mach number \citep{Mohapatra2019}. 


The second panel of \Cref{fig:time-evolution} shows that the amplitude of density fluctuations $\sigma_{s}$ increases with increasing strength of the stratification. For $\mathrm{Ri}\gtrsim0.05$, these buoyancy-induced density fluctuations are larger than the density fluctuations produced by unstratified turbulence, as seen in effectively unstratified turbulence runs with  $\mathrm{Ri}=0$ and $0.003$. 
Column 7 of \Cref{tab:sim_params} lists $\sigma_s$ for different runs. We discuss more about $\sigma_s$ and $\mathrm{Ri}$ in \cref{subsec:sig-Ri}. 

The vertical motions of parcels of gas also convert $z$-direction kinetic energy into gravitational potential energy due to work done against the buoyancy force. This mechanism of energy transfer is absent in unstratified turbulence.
The third and fourth panels of \Cref{fig:time-evolution} show that $E_{u_b}$ increases and $E_{u_z}$ decreases and the ratio $E_{u_b}/E_{u_z}$ increases with increasing stratification, as more and more of the driven energy is converted into potential energy. An increasing potential energy with increasing $\mathrm{Ri}$ is expected from a simple model of a driven-damped harmonic oscillator.


\subsection{Volume and column density structure}\label{subsec:columndens}
Here we present mock observables such as column density (similar to X-ray surface brightness if density fluctuations are small as in the present case) and the relative fluctuations thereof. We also plot density slices to look into the effect of integrating along the line of sight (LOS) and superimpose the corresponding velocity field. 

\begin{figure*}
	\centering
	\includegraphics[width=2.\columnwidth]{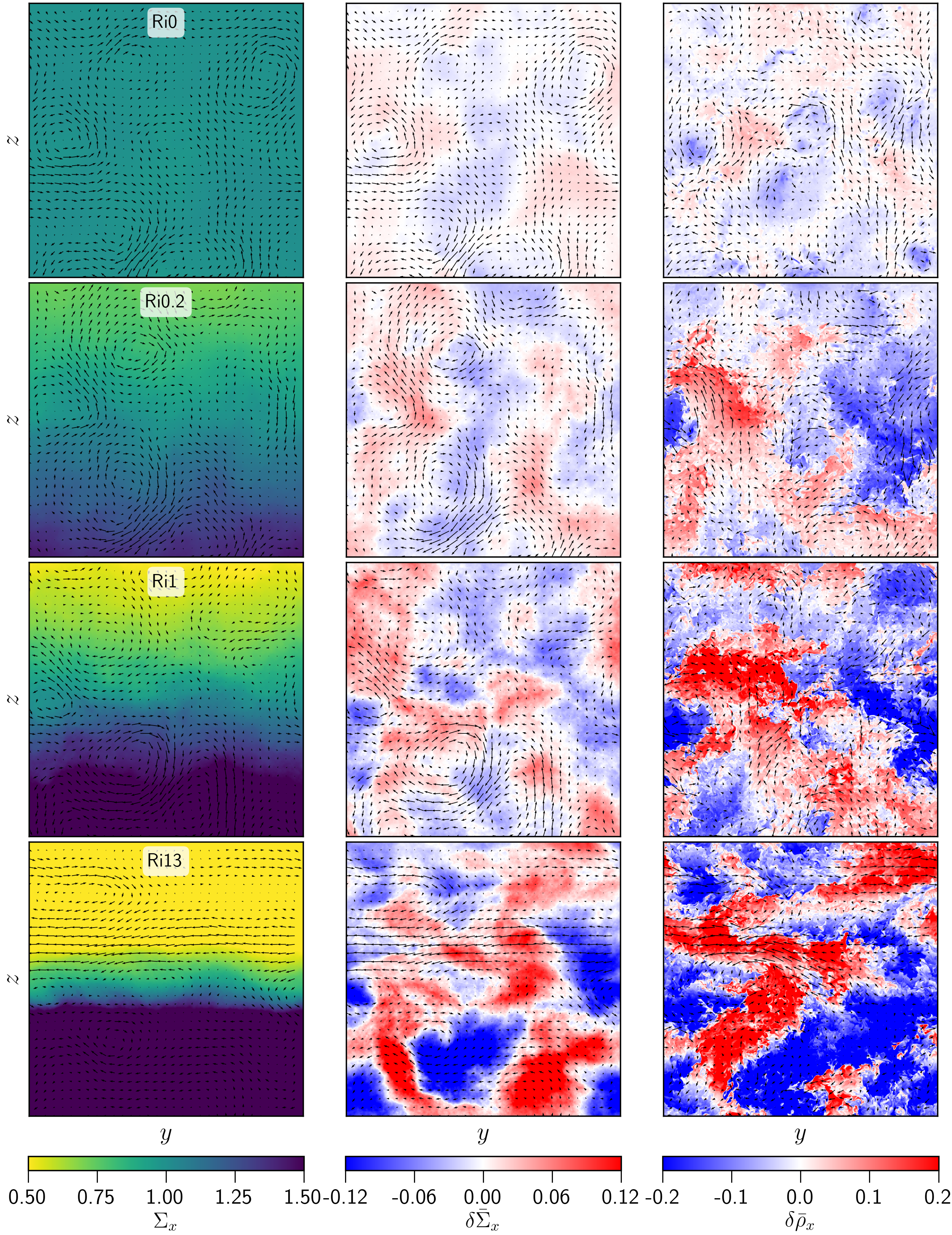}	
	\caption[Density projection along at $x$]{\underline{First column:} $\Sigma_x$, the density projection along $x$ (perpendicular to the stratification direction), for runs Ri0, Ri0.2, Ri1 and Ri13 at $t=6\ t_{\mathrm{turb}}$. The arrows indicate the integrated velocity field in the $\mathrm{yz}$ plane. The density gradient along $z$ is stronger for stronger stratification. \underline{Second column:} Corresponding $\bar{\Sigma}_x$ perturbations $\delta \bar{\Sigma}_x$ for different runs. These are obtained by dividing the observed density snapshot by the average density profile for that time and subtracting $1$ from it. Column density perturbations are significantly stronger for stronger stratification runs. \underline{Third column:} Normalised density perturbation slices $\delta \bar{\rho}_x$ taken in $x=0$ plane. These plots show more small-scale features and are of larger amplitude than $\delta \bar{\Sigma}_x$.}\label{fig:nrho-proj}
\end{figure*}

The left panel of \Cref{fig:nrho-proj} shows the density and velocity fields integrated perpendicular to the stratification direction (along $x$ direction, for simulations Ri0, Ri0.2, Ri1, and Ri13 (from top to bottom), i.e., from pure turbulence (top) to strong stratification (bottom), at $t=6\,t_\mathrm{turb}$. We denote column density and column density fluctuations (normalised with respect to the stratification profile) by $\Sigma_i$ and $\delta\bar{\Sigma}_i$ respectively, where
$\Sigma_i=\int\rho \mathrm{d}i$ and $\delta\bar{\Sigma}_i=\int\bar{\rho} \mathrm{d}i-1$, with $i$ denoting the line of sight (LOS) direction. These $\Sigma_i$ and $\delta\bar{\Sigma}_i$ plots provide a sense of comparison between the 3D structure of the ICM and the X-ray surface brightness observations. For $x$ as the LOS direction, the surface brightness ($\mathrm{SB}$) of ICM gas is proportional to $\int \rho^2 \mathrm{d}x$. The surface brightness fluctuations $\delta\mathrm{\bar{SB}}$ for small density fluctuations ($\delta \rho/\rho_0 < 1$) are then  proportional to  $\int2 \rho_0^2 \delta \bar{\rho} \mathrm{d}x\approx 2 \rho_0^2 \delta\bar{\Sigma}_x$.

We look for possible correlations between these mock observables and their dependence on $\mathrm{Ri}$. From the middle panel of \Cref{fig:nrho-proj}, we find that $\delta\bar{\Sigma}_x$  increases with increasing $\mathrm{Ri}$. This is what we expect, since $\sigma_s$ increases with $\mathrm{Ri}$ (second panel, \Cref{fig:time-evolution}).
 
Density slices (normalised with respect to the stratification profile) parallel to the $i$-axis passing through the origin are denoted by $\bar{\rho}_i$ and perturbation slices by $\delta\bar{\rho}_i$ ($\bar{\rho}_i=\bar{\rho}\vert_{i=i}$, $\delta\bar{\rho}_i=(\bar{\rho}-1)\vert_{i=i}$). The density fluctuations $\delta\bar{\rho}_i$ for different runs are plotted in the right panels of \Cref{fig:nrho-proj}.
We find that $\delta\bar{\rho}_x$ also increases with increasing $\mathrm{Ri}$. We see more small-scale structures in $\delta\bar{\rho}_x$ which disappear in $\delta\bar{\Sigma}_x$ due to integration along the $x$ direction. 

For $\mathrm{Ri}\gg1$ (bottom panel, \Cref{fig:nrho-proj}), the velocity field becomes primarily horizontal, the eddies become shorter in the vertical direction, and we observe layered stratified structures in $\delta\bar{\Sigma}_x$  and $\delta\bar{\rho}_x$.

\subsection{Velocity distribution}

\begin{figure}
	\includegraphics[width=\columnwidth]{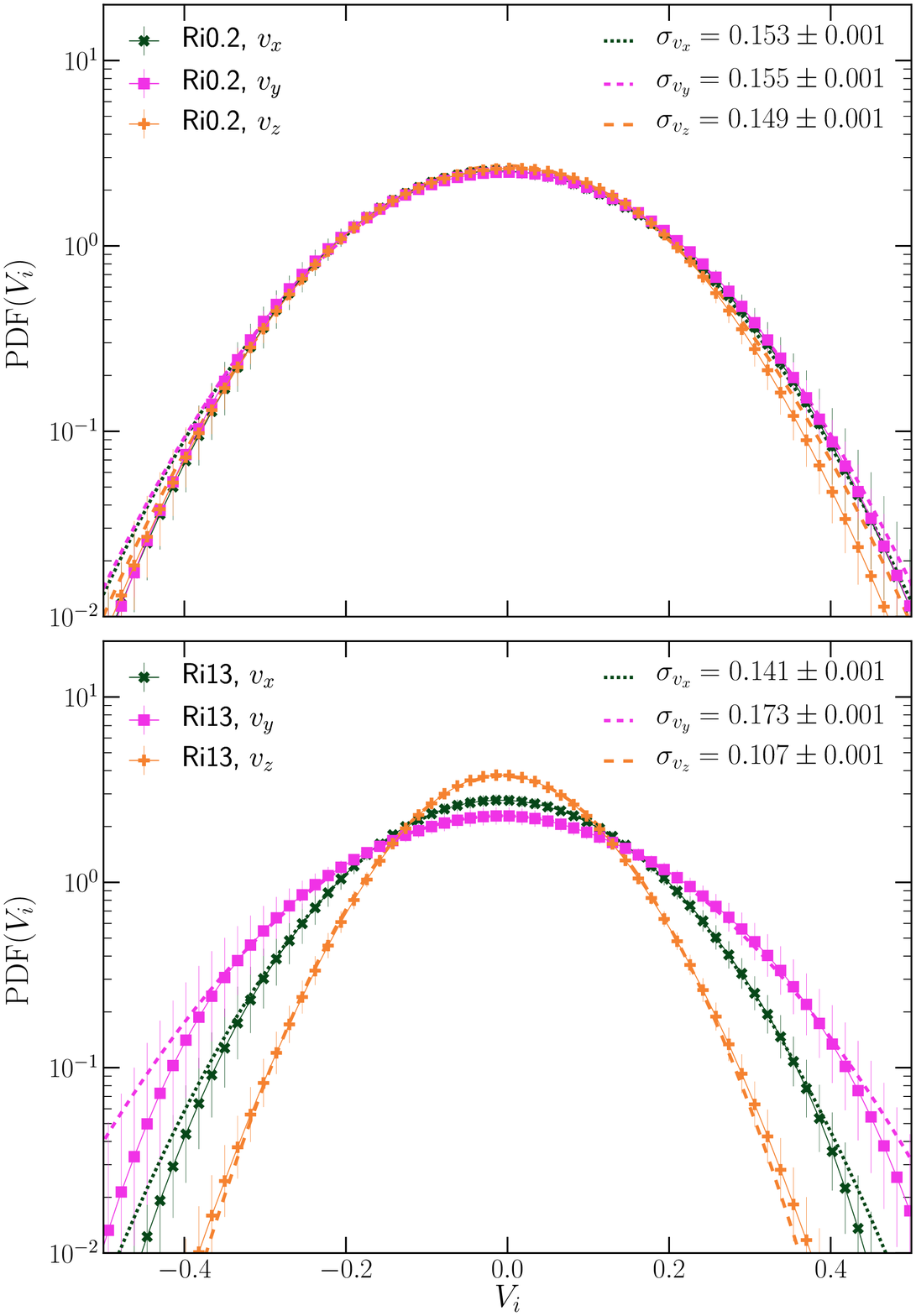}
	\hfill
	\caption [Velocity PDF]{Time-averaged volume-weighted velocity-component PDFs with Gaussian fits for $\mathrm{Ri}=0.2$ \emph{(top)} and $\mathrm{Ri}=13$ \emph{(bottom)}. The PDFs are Gaussian for all three components, but the velocity fluctuation is anisotropic for $\mathrm{Ri}\gtrsim1$.}\label{fig:velocity-pdf}
\end{figure}
The velocity field in normal unstratified turbulence is expected to follow a nearly Gaussian distribution, even for supersonic turbulence \citep[Figure A1 in ][]{Federrath2013}, and the velocity magnitudes of different components are supposed to be isotropic. Here we study the effect of stratification on the velocity distribution in different directions. 

In \Cref{fig:velocity-pdf}, we show the PDFs of the different velocity components for two different $\mathrm{Ri}=0.2$ and $\mathrm{Ri}=13$. Clearly, all the individual component PDFs for both the simulations are nearly Gaussian. For $\mathrm{Ri}=0.2$, all the three component-PDFs nearly overlap with each other and have the same standard deviation. From now on, we use $\sigma_{v_i}$ to denote the standard deviation of the velocity component along $i$. For $\mathrm{Ri}=0.2$, $\sigma_{v_z}$ is only slightly lower than $\sigma_{v_x}$ and $\sigma_{v_y}$ whereas for $\mathrm{Ri}=13$, the standard deviation of velocity in different directions is clearly anisotropic, with $\sigma_{v_z}<\sigma_{v_x}, \sigma_{v_y}$. 

This analysis shows that the velocity distribution is largely unaffected by the stratification if $\mathrm{Ri}\lesssim1$. However, for $\mathrm{Ri}\gtrsim1$, stratification strongly affects the velocity distribution by suppressing vertical motions. Also, strong gravity prevents turbulent mixing among different vertical layers (by flattening out the eddies along the direction of gravity), resulting in a small number of big two-dimensional eddies. This can lead to a difference between even $\sigma_{v_x}$ and $\sigma_{v_y}$ (see bottom panel in \Cref{fig:nrho-proj}) due to low number statistics of these eddies.


\subsection{PDF of density perturbations}\label{subsec:PDFdensperturb}
\begin{figure}
	\centering
	\includegraphics[width=0.99\columnwidth]{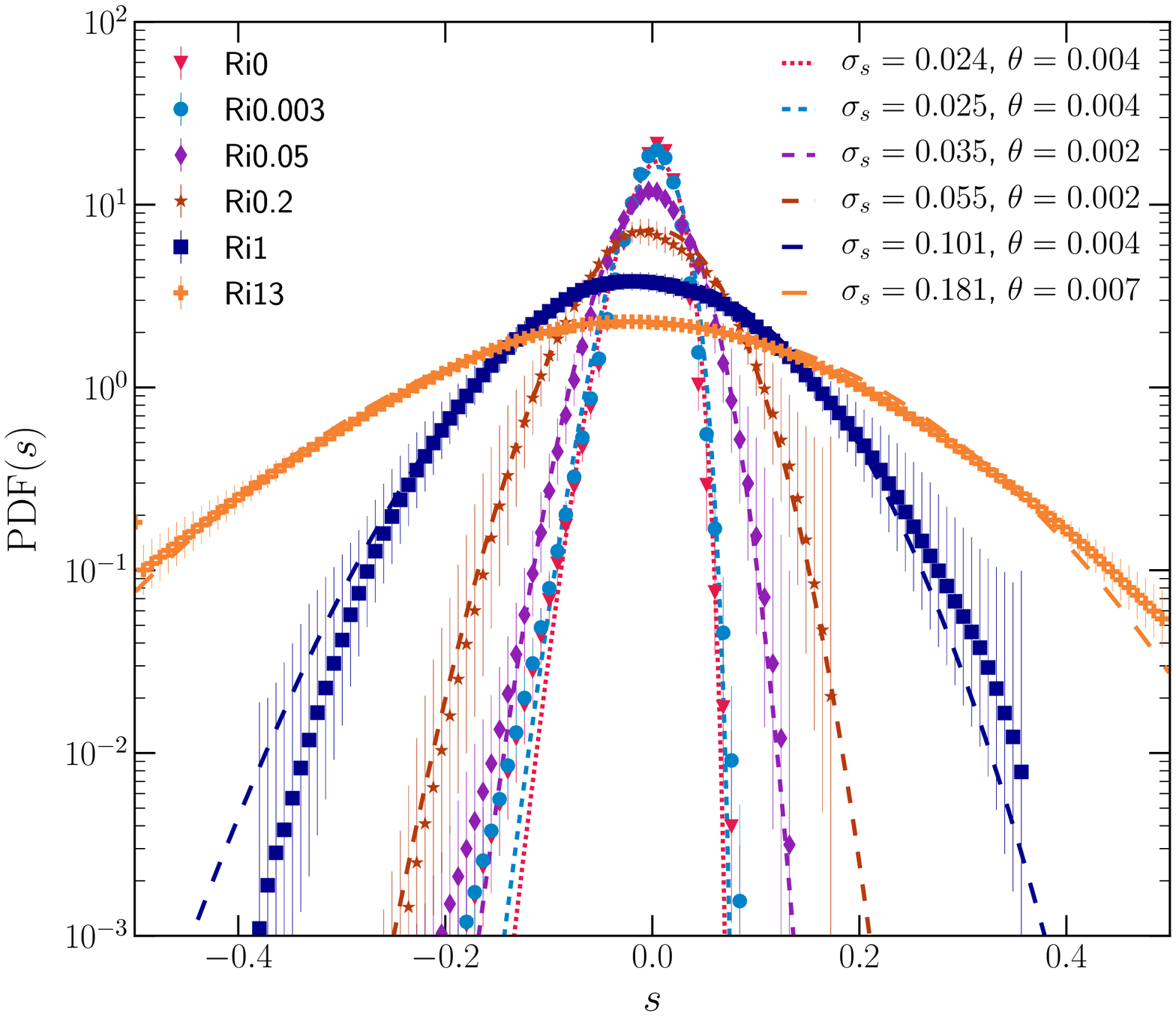}
	\caption[log-dens pdf]{Volume-weighted PDF of $s$, fit using \cref{eq:dens_hopkins}. The fluctuations (as characterised by the width of the PDF) are larger for stronger stratification. The intermittency parameter is close to $0$, as is expected for subsonic turbulence.}\label{fig:denshopkins}
\end{figure}
For unstratified subsonic turbulence, the density is supposed to follow a log-normal distribution \citep{Passot1998PhRvE, Federrath2008,zhuravleva2013quantifying,nolan2015} with some intermittency corrections \citep{Hopkins2013}. Since our density distribution will be affected by stratification, we instead show  the time-averaged  volume-weighted distribution of $s$ and fit the function $p_{\mathrm{HK}}(s)$ proposed by \cite{Hopkins2013}. It is defined as:
\begin{align}\label{eq:dens_hopkins}
	&p_{\mathrm{HK}}(s)=I_1\left(2\sqrt{\lambda w(s)}\right)\exp(-\lambda +w(s))\sqrt{\frac{\lambda}{\theta^2w(s)}},\\
	&\lambda\equiv\sigma_s^2/(2\theta^2),\  w(s)\equiv\lambda/(1+\theta)-s/\theta (w\geq 0), \nonumber
\end{align}
where $I_1(x)$ is the first-order modified Bessel function of the first kind. For further discussion on \cref{eq:dens_hopkins}, refer to \cite{Hopkins2013}.

 \Cref{fig:denshopkins} shows these fits with the $y$ axis in log scale for different stratification runs. Clearly, density fluctuations are larger for stronger stratification, corroborating the trends seen in \Cref{fig:time-evolution,fig:nrho-proj}.
 The density PDFs at lower levels of stratification ($\mathrm{Ri}\lesssim 0.1$) show a low-density tail, which is not well-accounted for by the fitting function.
 The value of the intermittency parameter $\theta$ is small for all different levels of stratification -- which implies that the distributions are close to log-normal. For $\theta=0$, Hopkins' non-log normal PDF $\approx$ log-normal PDF.  There is a lot of variation in the lower and higher density tails. We have confirmed that this variability is not affected by limited numerical resolution (\Cref{fig:PDF-convergence-120}). 
 For $\mathrm{Ri}\gtrsim1$, the PDFs do not show an asymmetric tail on the low-density side. This could be a result of density fluctuations being dominated by buoyancy oscillations instead of turbulent motions.

 The low-density tail is known to be a feature of the density PDFs when the adiabatic index $\gamma>1.$ \cite{Hopkins2013}' fit was shown to be a good fit for density PDF of gas following an isothermal equation of state, which has $\gamma=1$. We expect $\gamma$ to be a parameter of the distribution function \citep{nolan2015}.


\subsection{Density fluctuation $\sigma_s$ as a function of $\mathbf{Ri}$}\label{subsec:sig-Ri}

Here we study the effect of stratification, parameterised by $\mathrm{Ri}$, on the density fluctuation--Mach number ($\sigma_s-\mathrm{Ri}$) relation, and try to derive a theoretical scaling relation between the two. 
Other than its theoretical importance, these relations can be useful for estimating the velocity fluctuations from X-ray brightness observations of clusters. Once the rms density fluctuations are known from surface brightness maps, the velocity fluctuation can be calculated using these relations or similar relations between their spectra \citep{zhuravleva2013quantifying,zhuravleva2014relation}.

For our simulations with $\mathrm{Ri}\gtrsim 1$, the local $\mathrm{Ri}$ changes with $z$. Thus, we first create bins uniform in $\log_{10}(\mathrm{Ri})$. If for a particular simulation, the $\mathrm{Ri}$ value range falls in two different consecutive bin-ranges, we split the data at $z=z_0$, where $\mathrm{Ri}(z_0)=\mathrm{Ri}_{\mathrm{bin\  boundary}}$ between the two bins. These two datasets are cuboids with $-L/2\leq z <z_0$ and $z_0\leq z \leq L/2$. We then calculate $\sigma_s$ for these datasets separately. This method takes the variation of $\mathrm{Ri}$ as a function of $z$ within the simulation domain into account.

\begin{figure}
	\includegraphics[width=\columnwidth]{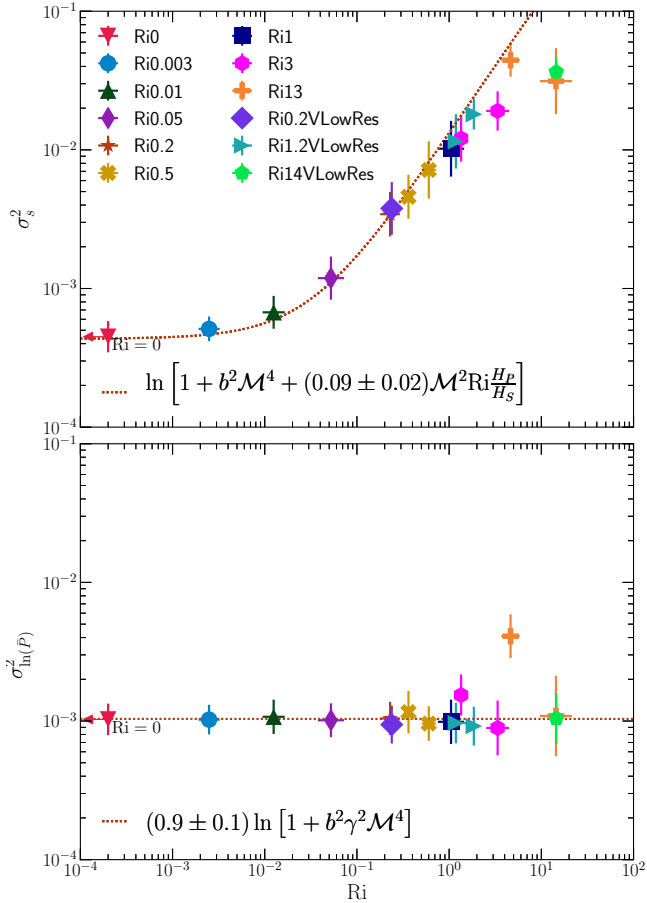}
	\hfill
	\caption [sigma-dens-Ri]{\underline{Upper panel:} Scatter plot of $\sigma_{s}$ versus $\mathrm{Ri}$. The amplitude of density fluctuations increases as a function of $\mathrm{Ri}$ for $0.01\lesssim \mathrm{Ri} <1$, reaches a peak at $ \mathrm{Ri} \approx 10$ and then starts decreasing for $\mathrm{Ri}>10$. \underline{Lower panel}: Scatter plot of $\sigma_{\ln(\bar{P})}$ (defined in \cref{subsec:sigP-Ri}) versus $\mathrm{Ri}$. Pressure fluctuations are roughly independent of $\mathrm{Ri}$. The outlier point has a larger local $\mathcal{M}$ which makes it consistent with the baseline relation. \underline{Note:} We conduct `VLowRes' runs with resolution $256^2\times384$, but with different $H_P/H_{\rho}$ (and thus different $H_P/H_S$), with simulation parameters described in \cref{tab:256_sim_params}. For these runs. we have plotted $\sigma_{s,\mathrm{scaled}}$ calculated using \cref{eq:sigma_scaled_up} instead of $\sigma_s$.}\label{fig:sig-Ri}
\end{figure}

The upper panel of \Cref{fig:sig-Ri} shows $\sigma_s$ as a function of $\mathrm{Ri}$. We see that $\sigma_s$ increases with $\mathrm{Ri}$ for $\mathrm{Ri}\gtrsim0.01$ and the increase seems to slow down around $\mathrm{Ri}\gtrsim1$. 
In the remaining part of this subsection, we attempt to derive a fitting function for this relation. We focus on the weakly stratified turbulence regime, with $0.01\lesssim\mathrm{Ri}\lesssim1$. 
The net density perturbations can be written as a sum of fluctuations due to unstratified turbulence and stratification, given by 
\begin{equation}\label{eq:Deltarhototal}
	\delta \bar{\rho}^2=\delta \bar{\rho}^{2}_{\mathrm{buoyancy}}+ \delta\bar{\rho}^{2}_{\mathrm{turb}}.
\end{equation}
 In \cite{Mohapatra2019}, we showed that $\delta\bar{\rho}^{2}_{\mathrm{turb}}\propto \mathcal{M}^4$. Here we attempt to derive an expression for $\delta \bar{\rho}^{2}_{\mathrm{buoyancy}}$. Substituting $N^2$ from \cref{eq:EoM} in \cref{eq:dens_buoy}, we obtain
\begin{equation}\label{eq:rho_buoy1}
	\mean{\delta \bar{\rho}^{2}_{\mathrm{buoyancy}}}= \frac{N^4}{g^2}\mean{\delta z^2}.
\end{equation}
We assume that the rms displacement of a parcel of gas in the turbulence dominated regime ($\mathrm{Ri}<1$) is proportional to the driving length scale. Thus, we substitute $\mean{\delta z^2}=\zeta^2L_{\mathrm{driv}}^2$, where $L_{\mathrm{driv}}$ is the turbulence driving length scale and $\zeta$ is a dimensionless constant. $N^2=\frac{g}{\gamma}\frac{\mathrm{d}S}{\mathrm{d}z}=\frac{g}{\gamma H_S}$. This gives us 
\begin{subequations}
\begin{align}
		\mean{\delta \bar{\rho}^{2}_{\mathrm{buoyancy}}}&= \frac{\zeta^2L_{\mathrm{driv}}^2}{\gamma^2H_S^2}\label{eq:disp_calc}\\
		&=\frac{\zeta^2 \mathrm{Ri}\mathcal{M}^2 c_s^2}{N^2\gamma^2H_S^2} \text{(using \cref{eq:Ri})}\\
		&=  \zeta^2 \mathcal{M}^2\mathrm{Ri} \frac{H_P}{H_S} \text{(substituting $N$ from \cref{eq:EoM})}.\label{eq:dens_final_buoy}
\end{align}
\end{subequations}

Inspired by the density fluctuation-Mach number relations in \cite{nolan2015,Mohapatra2019} we combine the two rms values of density perturbations and propose a new relation:
\begin{equation}\label{eq:sig-mach-Ri-relation}
	\sigma_s^2= \ln(1+b^2\mathcal{M}^{4}+\zeta^2\mathcal{M}^2\mathrm{Ri}\frac{H_P}{H_S}),
\end{equation}
where $b$ stands for the turbulence driving parameter and $b=1/3$ for solenoidal forcing \citep{Federrath2008,federrath2010}. We use $\zeta^2$ in \cref{eq:sig-mach-Ri-relation} as a fitting parameter for \Cref{fig:sig-Ri} and obtain $\zeta^2=0.09\pm0.02$. 

For $\mathrm{Ri}=0$ (unstratified turbulence), this reduces to $\sigma_s^2 =\ln(1+b^2\mathcal{M}^{4})$, which reproduces the $\sigma_{\bar{\rho}}\propto\mathcal{M}^2$ scaling relation in \cite{Mohapatra2019}. This relation for unstratified turbulence was motivated for subsonic flows by assuming the flow to be close to incompressible. In such a case, taking the divergence of \cref{eq:Euler2} gives the Poisson equation $\nabla^2 P = \nabla \cdot(\nabla \cdot (\rho \mathbf{v}\otimes\mathbf{v})) = \rho (\nabla {\bf v})^{\mathrm{T}} : \nabla {\bf v}$, which on simplifying gives $\delta P \sim \rho \delta v^2$, or 
$\delta \bar{P} \sim \gamma \delta v^2/c_s^2 \sim \gamma \mathcal{M}^2$. Now since these fluctuations are adiabatic \citep{Mohapatra2019}, $\delta \bar{\rho}=\delta \bar{P}/\gamma\propto \mathcal{M}^2$. 
Comparing this relation to the gamma-dependent $\sigma_s-\mathcal{M}$ relation of \cite{nolan2015} for $\gamma=5/3$, who obtain $\sigma_s^2=\ln(1+b^2\mathcal{M}^{28/9})\approx\ln(1+b^2\mathcal{M}^{3.1})$, which is shallower compared to our proposed relation. This discrepancy arises because our simulations are subsonic while their fits are obtained for the transonic regime. A flattening of $\sigma_s$ versus $\mathcal{M}$ for $\mathcal{M} \sim 1$ is also seen in Figure 1 of \citet{Mohapatra2019}.

Taking the ratio of the two types of density fluctuations, we obtain 
\begin{equation}
	\frac{\delta \bar{\rho}^{2}_{\mathrm{buoyancy}}}{\delta\bar{\rho}^{2}_{\mathrm{turb}}}=\frac{0.09\mathrm{Ri}}{b^2M^2}\frac{H_P}{H_S}\approx 30 \ \mathrm{Ri},
\end{equation}
inserting all other parameters which are constant for our simulations. Thus the effects of stratification should start dominating for $\mathrm{Ri}\gtrsim0.03$, which is exactly what we see in our simulations.

For $\mathrm{Ri}\gtrsim1$, the eddies start becoming flatter in the $z$ direction (see \cref{fig:nrho-proj}, fourth row panels), so $\zeta$ should depend on $\mathrm{Ri}$. We expect $\zeta$ to decrease with increasing $\mathrm{Ri}$ but we have not obtained a functional form for its dependence. Also for simulations with $\mathrm{Ri}\gtrsim 1$, both $\mathrm{Ri}$ and $\mathcal{M}$ show large variations within the simulation domain as a function of $z$. Obtaining an accurate and reliable $\sigma_s-\mathcal{M}-\mathrm{Ri}$ relation for $\mathrm{Ri}\gtrsim1$ is beyond the scope of the current study.

In order to check the dependence of $\sigma_s$ on $H_P/H_S$, we have run three simulations with grid resolution $256^2\times384$ and different values of this parameter (\Cref{tab:256_sim_params}). We use a constant value of $H_P/H_S$ (corresponding to $H_P/H_\rho=2$) in fitting $\sigma_s$--$\mathcal{M}$--$\mathrm{Ri}$ relation  shown in the upper panel of \cref{fig:sig-Ri}. To compare with our fit based on high-resolution runs with $H_P/H_\rho=2$, we scale $\sigma_s$ values from these simulations to $\sigma_{s,\mathrm{scaled}}$, which accounts for a different $H_P/H_S$ in \Cref{eq:sig-mach-Ri-relation}. The scaled $\sigma_s$ is given by 
\begin{align}\label{eq:sigma_scaled_up}
    &\sigma_{s\mathrm{,scaled}}=\ln\left[1+b^2\mathcal{M}^4+f_{\mathrm{scale}}\left(\exp(\sigma_s)-\left(1+b^2\mathcal{M}^4\right)\right)\right]\text{,}\\
    &\text{where }
    f_{\mathrm{scale}}=\frac{H_P/H_S|_{H_P/H_\rho=2}}{H_P/H_S|_{H_P/H_\rho}}.\nonumber
\end{align} 
This scaling is equivalent to multiplying $\delta \bar{\rho}_{\rm buoyancy}^2$ with $f_{\mathrm{scale}}$ (see \Cref{eq:dens_final_buoy}).

\cite{Shi2019} study the decay of a turbulent velocity field in a stratified medium and find that normalised density dispersion $\sigma_{\bar{\rho}}$ is smaller than the saturated value initially, stays roughly constant, and saturates after $t\gtrsim 1/N$ to $\sigma_{\bar{\rho}}=0.59 \mathcal{M}$ as turbulence decays with time and $\mathcal{M}$ decreases (see their figure 8).
Their temporal behaviour is consistent with our $\sigma_s $--$\mathcal{M}$--$\mathrm{Ri}$ relation, since the density fluctuations due to buoyancy with moderate stratification only depend on the driving scale and the entropy scale height (Eq.~\ref{eq:disp_calc}). These are held fixed while $\mathcal{M}$ decreases and $\mathrm{Ri}$ increases with time in decaying turbulence. At late times $\mathrm{Ri} \gtrsim 10$, $\sigma_s$ decreases linearly with decreasing $\mathcal{M}$ (fig.~8 in \citealt{Shi2019}) as we enter the strongly stratified regime, and vertical displacement and $\sigma_s$ are suppressed. This regime will be investigated further in a follow-up study.

\Cref{eq:disp_calc} shows that the net density perturbations depend only on the driving length scale of turbulence and not on the velocity itself - this would have significant implications for obtaining velocity from surface brightness fluctuations. We also find that $\sigma_s$  depends on three dimensionless parameters instead of just one - $\mathcal{M}$,  $\mathrm{Ri}$ and the ratio between the entropy and pressure scale heights, $H_S/H_P$. Thus, one cannot have a universal relation between density and velocity fluctuations.

\begin{table}
	\centering
	\caption{Simulation parameters for very low resolution runs}
	\label{tab:256_sim_params}
	\resizebox{0.45\textwidth}{!}{
		\begin{tabular}{lccccc} 
			\hline
			Label & $\mathrm{Ri}$ & $H_{\rho}$ & $H_P$ & $\sigma_s$ & $\sigma_{s\mathrm{,scaled}}$\\
			(1) & (2) & (3) & (4) & (5) & (6)\\
			\hline
			Ri0.2VLowRes & $0.2\pm0.1$ & $1.5$ &  $1.25$ & $0.03\pm0.01$ & $0.06\pm0.03$\\
			Ri1.2VLowRes & $1.2\pm0.4$ & $0.75$ &  $0.75$ & $0.06\pm0.03$ & $0.11\pm0.06$\\
			Ri14VLowRes & $14^{-8}_{+20}$ & $0.25$ &  $0.25$ & $0.11\pm0.03$ & $0.20\pm0.05$\\
		\hline
			
	\end{tabular}}
	\justifying \\ \begin{footnotesize} Notes: All these simulations have grid resolution $256^2\times384$ and $\mathcal{M}\approx0.25$. The columns (2) - (5) have their usual meanings. For these runs, $H_P/H_\rho\neq2$, unlike our runs in \Cref{tab:sim_params}. In column 6, we show $\sigma_{s,\mathrm{scaled}}$, which we calculate using \cref{eq:sigma_scaled_up}.\end{footnotesize} 
	
\end{table}

\subsection{Pressure fluctuation $\sigma_{\ln(\bar{P})}$ as a function of $\mathrm{Ri}$}\label{subsec:sigP-Ri}
We now shift our focus to to the relation between pressure fluctuations and Mach number, and its dependence on stratification. Pressure fluctuations can be inferred from the SZ observations  of clusters -- using this relation, we can estimate turbulent velocities on cluster outskirt scales ($\gtrsim 500 \mathrm{\ kpc}$) \citep{khatri2016,Simionescu2019SSRv,Mroczkowski2019}. A measurement of the turbulent velocity dispersion is required to calculate the level of non-thermal pressure support and thus 
the hydrostatic mass bias in clusters \citep{Cavaliere2011,Nelson2014ApJ}. 

In the lower panel of \cref{fig:sig-Ri}, we show $\sigma_{\ln(\bar{P})}$ as a function of $\mathrm{Ri}$. Here $\sigma_{\ln(\bar{P})}$ is the standard deviation of $\ln(\bar{P})$ distribution and is analogous to $\sigma_s$ for density. We find that $\sigma_{\ln(\bar{P})}$ is almost independent of the stratification. Among the two kinds of density fluctuations, $\delta\bar{\rho}_{\mathrm{buoyancy}}$ is isobaric so $\delta\bar{P}_{\mathrm{buoyancy}}=0$, whereas $\delta\bar{\rho}_{\mathrm{turb}}$ is adiabatic, so $\delta\bar{P}_{\mathrm{turb}}=\gamma\delta\bar{\rho}_{\mathrm{turb}}$.  Thus, the expression for the relation becomes:
\begin{equation}\label{eq:sigP_fluctuation}
	\sigma_{\ln(\bar{P})}=\ln(1+b^2\gamma^2\mathcal{M}^{4}),
\end{equation}
which is independent of $\mathrm{Ri}$. For $\mathrm{Ri}\gtrsim 1$, we see some variations in $\sigma_{\ln(\bar{P})}$. Since these simulations have large temperature gradients within the simulation domain (check \cref{tab:sim_params} for $H_P$ and $H_{\rho}$), the Mach number can vary as a function of $z$ even if we drive a homogeneous isotropic velocity field. This may cause $\sigma_{\ln(\bar{P})}$ to overshoot or undershoot.

While the $\sigma_s$--$\mathcal{M}$ fluctuation relation also depends on other parameters such as $H_P/H_S$ and $\mathrm{Ri}$, the $\sigma_{\ln(P)}$--$\mathcal{M}$ relation is independent of the stratification strength and the pressure and entropy scale heights. In \cite{Mohapatra2019}, we showed that the $\sigma_{\bar{\rho}}$--$\mathcal{M}$ relation depended on the thermodynamics (heating and cooling) whereas the $\sigma_{\bar{P}}$--$\mathcal{M}$ relation was independent of the thermodynamics, and still followed the same relation as homogeneous isotropic turbulence. Thus, for subsonic turbulence ($\mathcal{M}<1$), pressure fluctuations are independent of the thermodynamics and the stratification strength -- so they appear more reliable for estimating the velocity dispersion. This has important implications for different methods of measuring velocities and estimating the non-thermal pressure component in cluster outskirts. Higher resolution SZ observation of clusters in  future can also help us get velocities on smaller scales -- which are more relevant for understanding cool cores.

\subsection{Correlation between $v_z$ and density perturbations}\label{subsec:correlations}
\begin{figure}
	\includegraphics[width=\columnwidth]{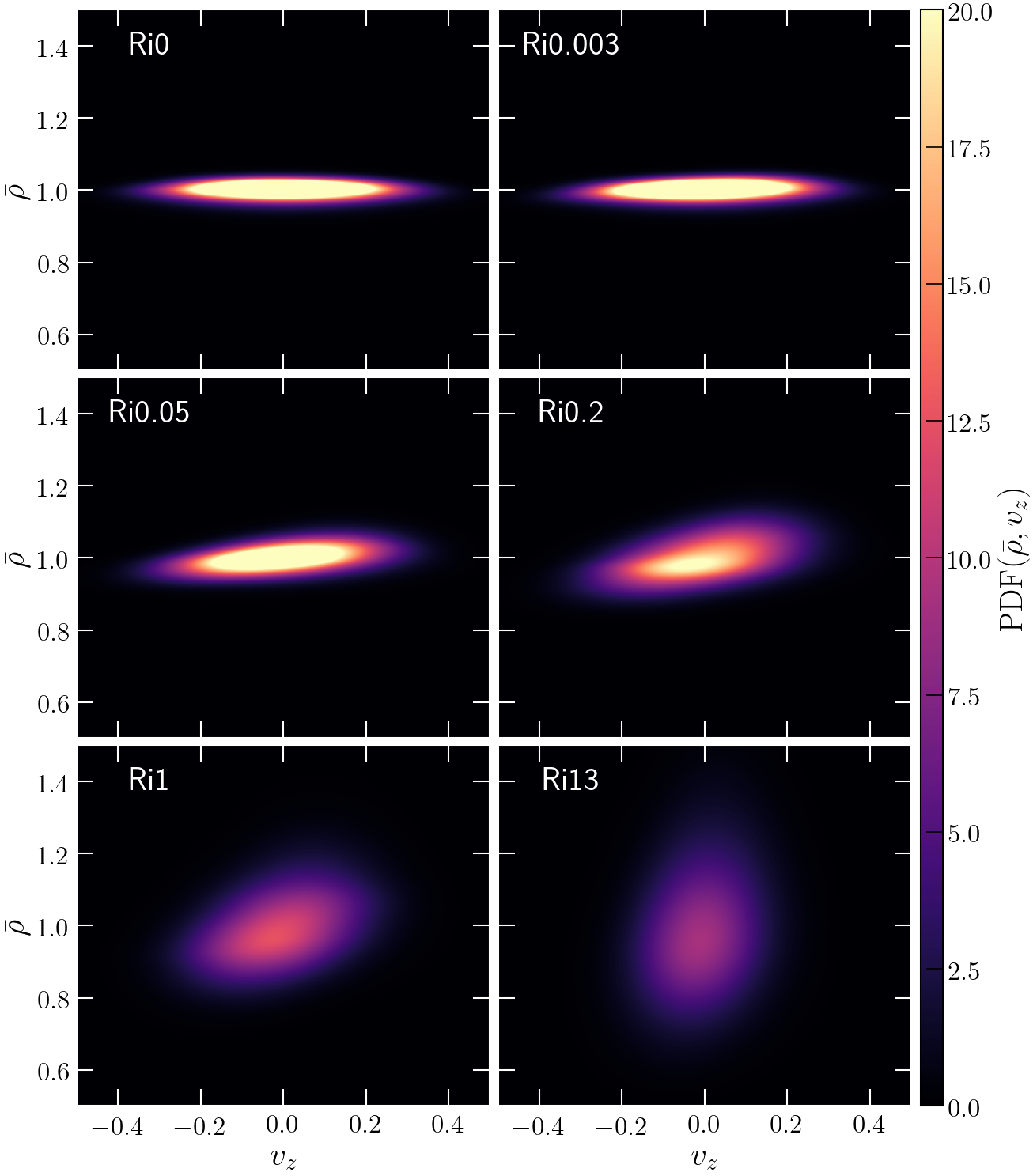}
	\hfill
	\caption [[Column density and Z velocity PDF]{Volume-weighted 2D PDF of $\bar{\rho}$ versus $v_z$. As the strength of the stratification increases ($10^{-3}<\mathrm{Ri}<1$) at constant $\mathcal{M}=0.25$}, the PDF tilts anticlockwise showing a local correlation between density and $v_z$. The correlation weakens for $\mathrm{Ri}\gtrsim1$.\label{fig:velz-nrho-pdf}
\end{figure}
In the run Ri0.2, one can observe a correlation between the signs of $v_z$ and the column density perturbations (second row from top, second and third columns, \Cref{fig:nrho-proj}). The red regions ($\delta \bar{\Sigma}_x>0$ and $\delta \bar{\rho}_x>0$) are more populated with upward facing arrows ($v_z>0$) and the blue regions ($\delta \bar{\Sigma}_x<0$ and $\delta \bar{\rho}_x<0$) are more populated with downward facing arrows ($v_z<0$). To look more into this,  in \Cref{fig:velz-nrho-pdf}we show two-dimensional volume-weighted probability distribution functions (2D PDFs) of normalised density fluctuations ($\delta\bar{\rho}$) and $v_z$. For unstratified turbulence ($\mathrm{Ri}=0$), $\bar{\rho}$ and $v_z$ are not locally correlated since turbulence is homogeneous and isotropic. As the stratification strength increases, the PDF starts rotating anticlockwise, showing a positive correlation between the direction of $v_z$ and the sign of $\delta \bar{\rho}$. However, for the strongest stratification runs, this correlation disappears. The spread in $v_z$ becomes narrower with increasing $\mathrm{Ri}$ and the spread in $\bar{\rho}$ increases.


A positive correlation between density fluctuations and vertical velocities is a consequence of the work done against the buoyancy force which converts kinetic energy into potential energy. It is a hallmark of stably stratified turbulence \citep{verma2018}.
For very strong stratification with $\mathrm{Ri}>10$, the $\delta\bar{\rho}-v_z$ correlation disappears, since now BV oscillations are generated by turbulent motions. This second transition is seen in the last two panels of \Cref{fig:velz-nrho-pdf}. Further, turbulent motions along the $z$ direction are heavily suppressed due to strong stratification. When we catch a parcel of gas with $\delta z>0$ and 
$\delta \rho>0$, we are equally likely to catch it in the rising and falling parts of its oscillatory motion and 
$v_{\mathrm{z, BV}}$ and $\delta z$ are not correlated. This is reminiscent of the phase plot (a plot of position versus velocity) of a simple harmonic oscillator.

\subsection{Nature of density perturbations}\label{subsec:prs-rho-PDF}
\begin{figure}
	\includegraphics[width=\columnwidth]{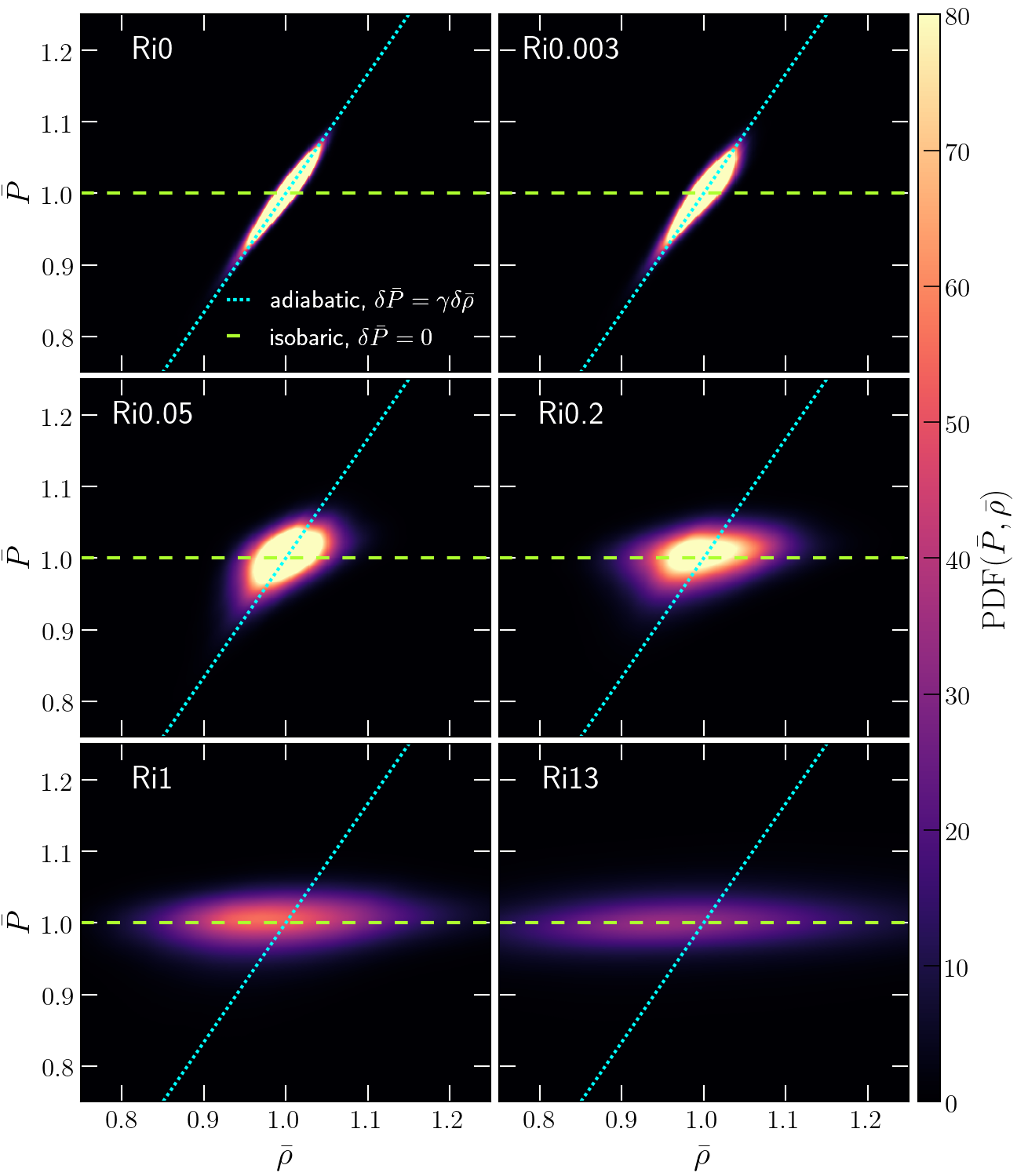}
	\hfill
	\caption [[Pressure and density PDF]{Volume-weighted 2D PDF of $\bar{P}$ versus $\bar{\rho}$ for different stratification runs ($10^{-3}<\mathrm{Ri}<1$, constant $\mathcal{M}=0.25$), with fits showing the nature of the perturbations. For the adiabatic fit, $\delta P/\mean{P} = \gamma\delta\rho/\mean{\rho}$, and for the isobaric fit, $\delta P/\mean{P}=0$. Fluctuations are isobaric for strong stratification and become increasingly adiabatic with the decreasing strength of the stratification.}\label{fig:pres-dens-pdf}
\end{figure}
X-ray observations of clusters have been used to characterise brightness fluctuations (caused by density fluctuations) according to their equation of state \citep{Arevalo2016ApJ, Churazov2016MNRAS, zhuravleva2018}. By comparing the emission in hard and soft X-ray bands, \citeauthor{zhuravleva2018} were able to categorise these fluctuations into adiabatic, isobaric or isothermal density fluctuations. 

We aim to distinguish between the two kinds of density perturbations in our simulations - the first caused by unstratified turbulence ($\delta \rho_{\mathrm{turb}}$) and the second introduced due to buoyancy and stratification ($\delta \rho_{\mathrm{buoyancy}}$). We can do so by differentiating between the nature (the effective equation of state or EOS) of these perturbations. Without stratification,  $\delta \rho=\delta \rho_{\mathrm{turb}}$. In this case, we expect density fluctuations due to subsonic turbulence to be adiabatic ($\delta P/P = \gamma\delta \rho/\rho$)  \citep{Mohapatra2019}. This is simply a consequence of a fluid element conserving its entropy over a time shorter than the turbulent turnover time. 

As we increase $\mathrm{Ri}$, we expect the contribution of $\delta \rho_{\mathrm{buoyancy}}$ to increase. To understand the nature of these buoyant density fluctuations, consider the parcel from \cref{subsec:CalcRi} again. When this parcel of gas rises, it is in constant pressure equilibrium with its surroundings (as its motion is subsonic). So $\delta P=0$ for the parcel in its new environment. The movement of the parcel only causes an overall change in its density with respect to its immediate surroundings. Hence density fluctuations caused by buoyancy should be effectively isobaric in nature for subsonic velocities ($\mathcal{M}<1$).

To confirm our two hypotheses, we show 2D PDFs of $\bar{P}$ and $\bar{\rho}$ and look for possible correlations between them (\Cref{fig:pres-dens-pdf}). The two dashed lines indicate the relations between density and pressure if the fluctuations were adiabatic ($\delta\bar{P}=\gamma\delta\bar{\rho}$) and isobaric ($\delta\bar{P}=0$), respectively. The total density fluctuations can be given by \cref{eq:Deltarhototal}. As expected, the PDF for unstratified turbulence closely follows the adiabatic EOS fit with little spread. Further on increasing stratification, the PDF starts spreading out horizontally due to the increasing contribution of isobaric density fluctuations. This is because the contribution of $\delta \rho_{\mathrm{buoyancy}}$  to $\delta \rho$ increases, in agreement with what we expected.

\cite{zhuravleva2018} find that for most clusters, the density fluctuations in the inner half of cool-cores are either isothermal or isobaric. We propose that density fluctuations caused by turbulent motion of gas in a stratified medium can be a key contributor to the isobaric density fluctuations. Thermal conduction can make density fluctuations isothermal at small scales, but a detailed investigation of this is beyond the scope of the present study.

\subsection{Power spectra}\label{subsec:densvelpow}
While all the 1D and 2D PDFs tell us about different global statistics of stratified turbulence, we are also interested in the spatial scaling of density and velocity perturbation fields. One way to study the scale dependence is by computing power spectra of these fields. Recent observational works such as \cite{zhuravleva2014turbulent,zhuravleva2018} have used power spectra to calculate density and velocity fluctuations of ICM gas as a function of scale. Theoretical studies such as \citet{zhuravleva2013quantifying} and \citet{gaspari2014} model these relations between density and velocity power spectra in cosmological simulations and large-scale cluster simulations, respectively. However, in \Cref{fig:cluster-Ri-profile} we showed that $\mathrm{Ri}$ can vary significantly with radial distance from the cluster centre. Hence we need to study the power spectra on local scales, so that we can study the dependency (if any) of these power spectra on $\mathrm{Ri}$. Then we can use these models to relate density and velocity spectra based on $\mathrm{Ri}$ as a function of radius in the ICM.

By following different binning methods in $k$-space, we can also study the effect of stratification-induced anisotropy on the distribution of power in different directions (perpendicular and parallel to the stratification direction). 

\subsubsection{Velocity power spectra}\label{subsubsec:velspow}
\begin{figure}
	\includegraphics[width=\columnwidth]{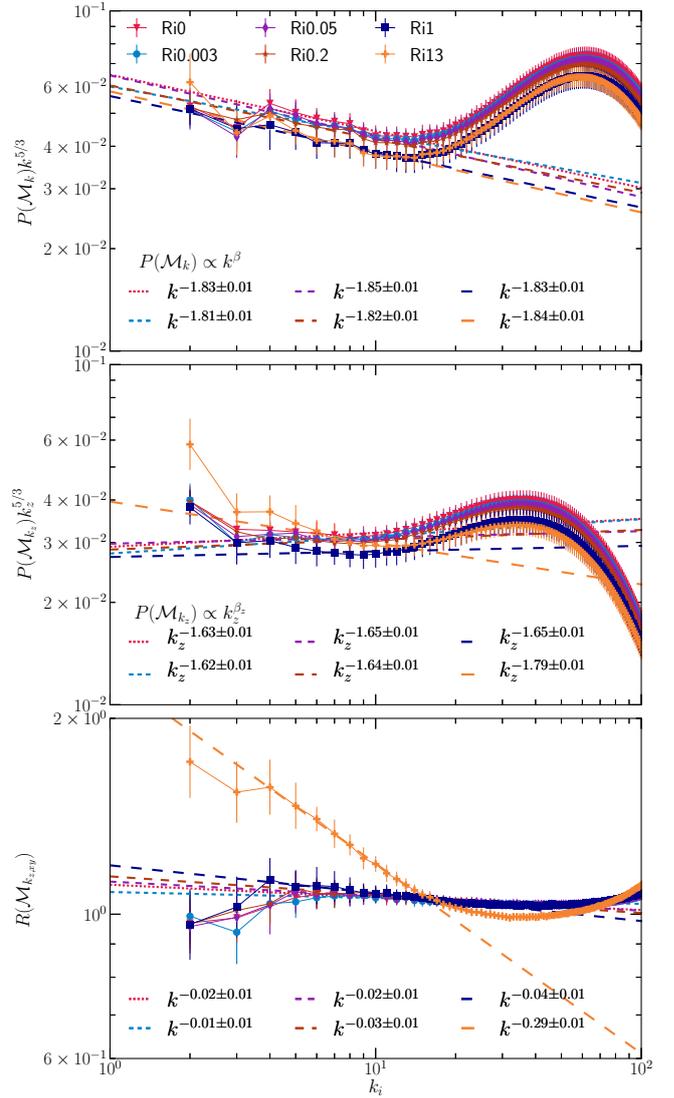}
	\hfill
	\caption [Mach number power spectrum]{\underline{Upper panel:} Time-averaged compensated power spectra of Mach number (velocity fluctuations normalised to the sound speed) for our standard set of runs with $\mathrm{Ri}$ indicated in labels. The slope of the power spectrum in the inertial range is steeper than the Kolmogorov scaling, but does not vary much with $\mathrm{Ri}$. \underline{Middle panel}: The same as the top panel, but with binning along $k_z$. \underline{Lower panel:} Ratio of Mach number power spectra in vertical ($z$) and horizontal ($x,y$) modes, $R(\mathcal{M}_{k_{z,xy}})=P(\mathcal{M}_{k_z})/(0.5(P(\mathcal{M}_{k_x})+P(\mathcal{M}_{k_y}))$. \underline{Note:} All fit slopes indicated in legends are fits for the corresponding power spectra and not for compensated power spectra.}\label{fig:mach-pow-spectrum}
\end{figure}
We show the power spectrum of the local Mach number (velocity in units of the local sound speed) $P(\mathcal{M}_k)$ in the top panel of  \Cref{fig:mach-pow-spectrum}.  To evaluate these spectra, we bin power in Fourier space in spherical shells of width $\delta k=1$ centred at $\mathbf{k}=0$. We compensate $P(\mathcal{M}_k)$  by the \cite{kolmogorov1941dissipation} scaling ($k^{-5/3}$) of velocities for homogeneous isotropic incompressible turbulence. $P(\mathcal{M}_k)$ neither changes in slope nor in amplitude with changing strengths of stratification. The constant amplitude is expected since $\mathcal{M}\approx0.25$ for all runs. We find slopes of $P(\mathcal{M}_k)$ in the range $-1.81\pm0.01$ to $-1.86\pm0.01$, without any obvious systematic dependence on Ri. The spectrum 
is steeper than the Kolmogorov scaling as is expected from intermittency effects \citep{She1994,Boldyrev2002,Schmidt2008,konstandin2012,Federrath2013}. \cite{She1994} predict a steepening of the power spectrum by $\approx0.03$ for incompressible turbulence. However, here we are dealing with mildly compressible turbulence ($\mathcal{M}\sim0.25$), which leads to additional steepening of the velocity power spectrum \citep{Galtier2011PhRvL,Aluie2011PhRvL,Banerjee2013PhRvE,Banerjee2014JFM,Federrath2013}. 

In order to understand the trends in the spectra and to look at the distribution of power perpendicular and parallel to the stratification direction, we bin the squared Fourier amplitudes separately  along slabs of constant $k_i$, with $\delta k_i=1$, centered at $k_i=0$, where $i$ can be $x$, $y$ or $z$. We denote these as $P(\mathcal{M}_{k_i})$. We also compute the ratio of power parallel and perpendicular to the stratification direction, denoted by $R(\mathcal{M}_{k_{z,xy}})$, where $R(\mathcal{M}_{k_{z,xy}})=P(\mathcal{M}_{k_z})/(0.5(P(\mathcal{M}_{k_x})+P(\mathcal{M}_{k_y}))$.
We show  $P(\mathcal{M}_{k_z})$ in the middle panel of \Cref{fig:mach-pow-spectrum}, compensated by $k_z^{-5/3}$. In the lower panel, we show $R(\mathcal{M}_{k_{z,xy}})$. One can infer the scaling perpendicular to $k_z$ and the amplitude of velocity fluctuations by dividing the slopes of the middle panel by those in the lower panel. 

From the middle panel of \Cref{fig:mach-pow-spectrum}, we observe that while the amplitude of the power spectrum remains almost the same, the slope of $P(\mathcal{M}_{k_z})$ steepens around $\mathrm{Ri}\gtrsim1$. This is due to anisotropy effects - the eddies in strongly stratified turbulence are flattened (like pancakes) parallel to the direction of gravity (see lower panel of \Cref{fig:nrho-proj}). Thus, $k_z\gg k_x,k_y$ and the power spectrum $P(\mathcal{M}_{k_z})$ becomes steeper with increasing $\mathrm{Ri}$ for $\mathrm{Ri}>1$. It can be clearly seen in the lower panel of \Cref{fig:mach-pow-spectrum}, where $R(\mathcal{M}_{k_{z,xy}})$ is almost flat for $\mathrm{Ri}<1$ and steepens significantly for $\mathrm{Ri}\gtrsim1$, $k\lesssim20$. 

\subsubsection{Density power spectra}\label{subsubsec:denspow}
\begin{figure}
	\includegraphics[width=\columnwidth]{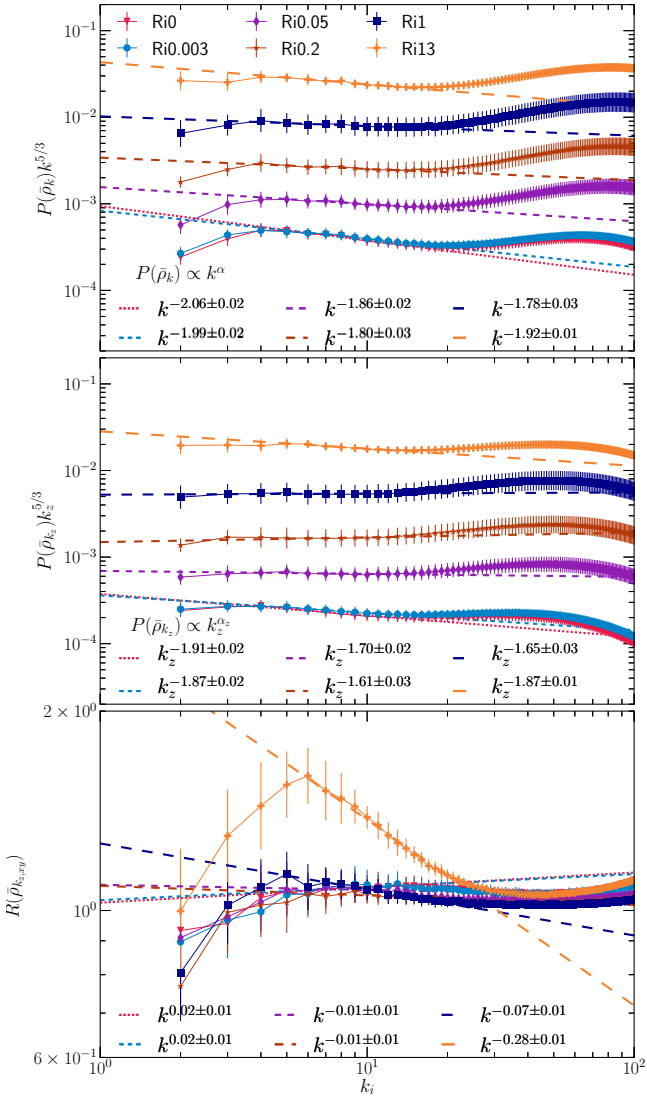}
	\hfill
	\caption [Density power spectrum]{\underline{Upper panel:} Time-averaged compensated power spectra of density fluctuations for different runs with $\mathrm{Ri}$ indicated in the run labels. The slope of the power spectrum in the inertial range decreases as a function of $\mathrm{Ri}$. It becomes closer to \cite{corrsin1951spectrum} scaling at large $\mathrm{Ri}$. 
		\underline{Middle panel}: The same as the top panel, but with binning along $k_z$. \underline{Lower panel:} Ratio of normalised density power spectra in vertical ($z$) and horizontal ($x,y$) modes, $R(\bar{\rho}_{k_{z,xy}})=P(\bar{\rho}_{k_z})/(0.5(P(\bar{\rho}_{k_x})+P(\bar{\rho}_{k_y}))$. \underline{Note:} All fit slopes indicated in legends are fits for the corresponding power spectra and not for compensated power spectra.}\label{fig:dens-pow-spectrum}
\end{figure}
Now we study the effect of stratification and anisotropy on the amplitude of density perturbations over different scales. We do so by computing the power spectrum of $\bar{\rho}$ (compensated by $k^{-5/3}$ \citealt{corrsin1951spectrum} scaling of passive scalars), in the top panel of \Cref{fig:dens-pow-spectrum} for different runs. As expected, the amplitude of the power spectrum increases with increasing stratification, by almost two orders of magnitude, as quantified in detail in \cref{subsec:sig-Ri}. The density power spectrum for unstratified turbulence overlaps with the weakly stratified run with $\mathrm{Ri}=0.003$. 

As for the velocity power spectra, we study the effects of anisotropy by computing $P(\bar{\rho}_{k_i})$, where $i=x,y\text{ or }z$, binned similar to $P(\mathcal{M}_{k_i})$ along $k_i$. We also compute the ratio of power parallel and perpendicular to the stratification direction, denoted by $R(\bar{\rho}_{k_{z,xy}})=P(\bar{\rho}_{k_z})/(0.5(P(\bar{\rho}_{k_x}))+P(\bar{\rho}_{k_y}))$.
We show $P(\bar{\rho}_{k_z})$ in the middle panel of \Cref{fig:dens-pow-spectrum}, compensated by $k_z^{5/3}$. In the lower panel, we show $R(\bar{\rho}_{k_{z,xy}})$. For $P(\bar{\rho}_{k_z})$, we see the same trend as for $P(\bar{\rho}_{k})$, in the sense that it increases in amplitude and the power spectrum first becomes shallower with increasing $\mathrm{Ri}$ and then starts becoming steeper again for $\mathrm{Ri}\gtrsim1$. For $R(\bar{\rho}_{k_{z,xy}})$, the slope remains close to zero for $\mathrm{Ri}<1$ and becomes steeper for $\mathrm{Ri}\gtrsim1$, which is very similar to what we observed for $R(\mathcal{M}_{k_{z,xy}})$. 

\begin{figure}
	\includegraphics[width=\columnwidth]{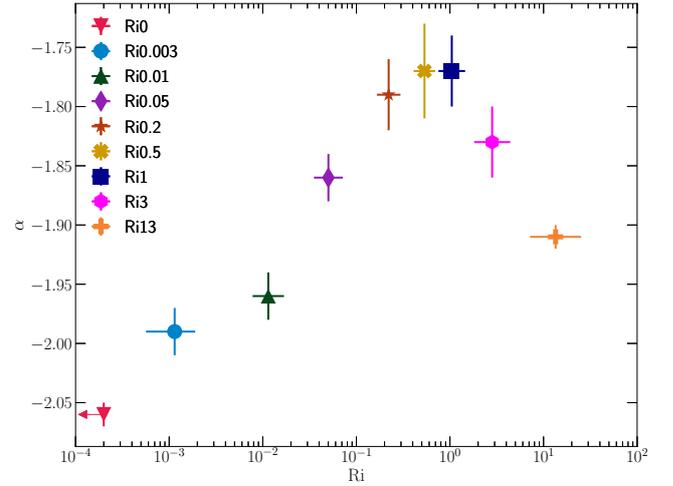}
	\hfill
	\caption [alpha-dens-Ri]{Scatter plot of $\alpha$, the slope of density power spectrum ($P(\bar{\rho}_k)\propto k^{\alpha}$). The slope becomes shallower with increasing $\mathrm{Ri}$ and peaks at $\mathrm{Ri}\sim1$ (\Cref{fig:mach-pow-spectrum}) before steepening again for $\mathrm{Ri}>1$.}\label{fig:alpha-Ri}
\end{figure}

We show $\alpha$, the spectral index of $P(\bar{\rho}_k)$ in \Cref{fig:alpha-Ri}. For low $\mathrm{Ri}\lesssim0.01$, $\alpha$ is more negative than the theoretically predicted \cite{corrsin1951spectrum} scaling ($k^{-5/3}$). With increasing stratification, the value of $\alpha$ initially starts to rise, peaks at $\mathrm{Ri}\sim1$ and then falls again for $\mathrm{Ri}>1$. 

The above trends reflect the two transitions we observe as we increase $\mathrm{Ri}$ from $\approx0.001$ to $\approx13$. The first transition occurs around $0.001<\mathrm{Ri}<0.01$, when $\delta\rho_{\mathrm{buoyancy}}$ starts dominating over $\delta\rho_{\mathrm{turb}}$. As seen in the power spectrum, unstratified turbulence by itself is unable to drive small-scale density perturbations very efficiently, due to which $P(\bar{\rho}_{k})$ is much steeper than the velocity power spectrum. However, with weak stratification, when the velocity distribution is still quite isotropic, $\delta\bar{\rho}$ is locally correlated to $v_z$, and the slope of the power spectrum becomes shallower and closer to the slope of the velocity power spectrum. On increasing stratification beyond $\mathrm{Ri}>1$, the eddies become more pancake-like in shape. This leads to a decrease in $v_z$, and $k_z>k_x,k_y$. Since the vertical motions of gas are still the dominant contributor to the total density fluctuations ($\delta\rho_{\mathrm{buoyancy}}>\delta\rho_{\mathrm{turb}}$ for $\mathrm{Ri}>0.03$), the overall power spectrum also becomes steeper. 
 
 One can calulate the slopes of the perpendicular density power spectrum $(P(\bar{\rho}_{k_{xy}})=0.5(P(\bar{\rho}_{k_x})+P(\bar{\rho}_{k_y}))$ by dividing the slope of the middle panel by the corresponding slopes of the lower panel. The slope of $P(\bar{\rho}_{k_{xy}})$ only becomes shallower till $\mathrm{Ri}\approx 1$ and then saturates at that value. This shows that strong stratification affects density modes only in the direction parallel to the stratification.

\cite{gaspari2013constraining} also derive the relation between gas density and velocity power spectra in their simulations. They find that normalised density fluctuations vary with the Mach number as $\delta\bar{\rho}\simeq1/4\mathcal{M}$. However, they use a fixed stratification profile for all their simulations. Their relation is very likely a result of strong stratification with $\mathrm{Ri} \gtrsim 10$. \Cref{fig:cluster-Ri-profile} shows the variation in $\mathrm{Ri}$ as a function of radius and \cref{eq:sig-mach-Ri-relation} is our density fluctuation--Mach number relation. Clearly, different terms in our relation will dominate in parts of the cluster. 

\subsubsection{Ratio between density and velocity power spectra}\label{subsubsec:etak}
\begin{figure}
	\includegraphics[width=\columnwidth]{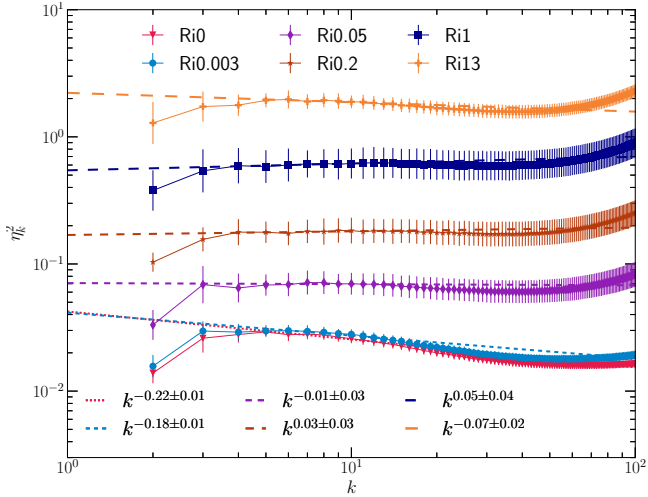}
	\hfill
	\caption [alpha-dens-Ri]{Ratio of density and velocity power spectra, $\eta_k^2$, for our standard set of runs. We see that $\eta_k$ increases with increasing $\mathrm{Ri}$.}\label{fig:etak}
\end{figure}

\cite{zhuravleva2014relation} derive the relation between density and 1D velocity Fourier amplitudes $\delta\rho_k$ and $v_{1,k}$, respectively. They show that the ratio of the normalised power spectra, given by\footnote{Note that, unlike here, in Fig. 16 of \citet{Mohapatra2019}, we define $\eta_k=(\delta\rho_k/\mean{\rho})/(v_k/c_s)$ using the 3D velocity power spectrum.}
\begin{equation}
    \eta_k= (\delta\rho_k/\mean{\rho})/(v_{1,k}/c_s)\approx \sqrt{3 P(\bar{\rho}_k)/P(\mathcal{M}_k)},
\end{equation}
is nearly constant with $\eta_k\approx1.0\pm0.3$ and independent of length scale or $\mathcal{M}$.

We show $\eta_k^2$ in \Cref{fig:etak} and find that it increases with the stratification strength, with $0.01\lesssim \eta_k^2\lesssim1$ for our simulations, and its amplitude increases with increasing $\mathrm{Ri}$. The ratio $\eta_k$ is roughly independent of $k$ only for $0.05\lesssim\mathrm{Ri}\lesssim2$, when the density and velocity power spectra are parallel due to the correlation we discussed in \cref{subsec:correlations}. Thus, most importantly, the conversion of density perturbations into velocity perturbations is not universal.

%
%

\section{Comparisons, caveats and future work}\label{sec:caveats-future}


\cite{Kumar2014} study the passive scalar power spectrum in stratified turbulence. Several other studies  \citep[e.g.,][]{Carnevale2001,Lindborg2006,Deusebio2013} discuss the transition between buoyancy and inertial length scale in the power spectra of velocity and passive scalars at the Bolgiano length scale $l_B$. In Bolgiano-Obukhov phenomelogy, for $\ell
>\ell_B$, buoyancy effects dominate the turbulent scaling relations and for $\ell<\ell_B$, turbulence dominates. Our simulations have enough resolution to resolve the Ozmidov length scale $\ell_O$ within the inertial range of turbulence (i.e., see the top panel of \Cref{fig:mach-pow-spectrum}; bottleneck effects dominate for $|k|\gtrsim20$). But as shown in \cite{Alam2019}, $\ell_B\ll \ell_O$, and is difficult to resolve in numerical simulations with our setup. Although we do not observe a scaling transition within the same power spectrum (probably due to a lack of resolution), we observe a transition in the power spectrum slope at around $\mathrm{Ri}\sim 1$, which corresponds to $\ell_O\approx L_{\text{driv}}$. This transition is even more evident in $P(\bar{\rho}_{k_z})$ and $P(\mathcal{M}_{k_z})$, and the corresponding ratios $R(\bar{\rho}_{k_{z,xy}})$ and $R(\mathcal{M}_{k_{z,xy}})$, which become steeper for $\mathrm{Ri}\gtrsim 1$.


There is a major difference between our simulations and those in the above studies -- the forcing field we use has components along all three directions (both perpendicular and parallel to the stratification profile), whereas the others have components only perpendicular to the profile. The difference in driving is mainly because the above studies model the stratified turbulence in the earth's atmosphere and oceans, while we are motivated by the turbulence in the ICM, where driving by AGN jets is not limited to directions perpendicular to the stratification profile. In order to test the dependence of our results on the driving, we conducted two simulations (Ri0.6NV and Ri10NV; see Tab.~\ref{tab:sim_params}) with turbulent driving only perpendicular to the stratification profile. We compare $\sigma_s$ and $\alpha$ (slope of density power spectrum) of these runs to corresponding runs Ri0.5 and Ri13 which have isotropic driving. We find that $\sigma_s$ and $\alpha$ for both these simulations lie within error bounds of each other (see \Cref{tab:sim_params} for $\alpha$ and $\sigma_s$). Even with 2D forcing, due to Kelvin-Helmholtz instability, the flow breaks down into 3D turbulence and similar results are obtained for 2D and 3D forcing. Even with horizontal driving, we see significant anisotropy in both velocity and density fields only for $\mathrm{Ri}\gtrsim 1$.

We propose a new relation between $\sigma_s$, $\mathcal{M}$, and $\mathrm{Ri}$, where density fluctuations depend on Mach number, Richardson number, and the ratio of entropy and pressure scale heights  ($\mathcal{M}$, $\mathrm{Ri}$ and $H_S/H_P$). But we have only exhaustively scanned the parameter space of $\mathrm{Ri}$. Following up on this, it is important to test this relation for different values of all the three dimensionless numbers. 
We also need to extend the fitting function to $\mathrm{Ri}\gtrsim 1$ (which is also relevant for clusters) and quantify the anisotropy in the velocity distribution.

For $\mathrm{Ri}\gtrsim1$ runs, we have large variations in $\mathrm{Ri}$ and temperature $T$ along the stratification direction. This prevents us from conducting a simulation for a particular $\mathrm{Ri}$ and having uniform $\mathcal{M}$ throughout the box.
For example, in our strongest stratification run Ri13, $\mathrm{Ri}$ increases from $3$ to $30$ and $\mathcal{M}$ decreases from $0.35$ to $0.15$ from $z=-0.5$ to $z=0.5$ (due to variation in $T$). $\sigma_s$, which depends on both these parameters, increases from $0.18$ at $z=-0.5$ to $0.24$ at $z\approx-0.2$, reaches a peak and then decreases with increasing $z$ to $0.1$ at $z=0.5$.
The setup also prevents us from simulating highly stratified turbulence with $\mathrm{Ri}\sim100$, which should be around the upper limit of the strength of stratification for clusters. 

Cooling and thermal instability, thermal conduction, and magnetic fields are some of the physics that are important for a realistic ICM and are not a part of this study. 
These effects can affect the density power spectrum, as seen in \cite{gaspari2014,Mohapatra2019}. While conduction tries to eliminate density fluctuations and gradients, stratification and cooling do quite the opposite. A parameter scan of different levels of cooling and thermal conduction and stratification will be an important study for the near future.

\section{Conclusions}\label{sec:Conclusion}
In this study, we have conducted idealised simulations of stratified turbulence with different levels of stratification using high-resolution hydrodynamic simulations. We have covered the parameter space $0.001\lesssim\mathrm{Ri}\lesssim10$. $\mathrm{Ri}\sim1$ is most relevant for modelling subsonic stratified turbulence in the ICM, which we have sampled with 9 simulations (\Cref{tab:sim_params}). The following are the main conclusions of our study:
\begin{itemize}
	\item The amplitude of density fluctuations (denoted in log-scale by $\sigma_s$) is dependent on the Richardson number ($\mathrm{Ri}$), and it increases with increasing $\mathrm{Ri}$ for $\mathrm{Ri}\lesssim10$. For weakly stratified subsonic turbulence (with $\mathrm{Ri}\lesssim1$, $\mathcal{M}<1$), we derive a new relation $\sigma_s^2 = \ln(1+b^2\mathcal{M}^{4}+0.09\mathcal{M}^2\mathrm{Ri}H_P/H_S)$. Thus, we find that the density fluctuations in a weakly stratified medium (the last term in the previous expression) depend on three dimensionless parameters: $\mathcal{M}$, $\mathrm{Ri}$ and $H_P/H_S$. For $\mathrm{Ri} \gtrsim 10$, the magnitude of density fluctuations may become independent of $\mathrm{Ri}$.
	\item The amplitude of pressure fluctuations is independent of $\mathrm{Ri}$ for $\mathrm{Ri}\lesssim 10$. For subsonic turbulence with $\mathrm{Ri}\lesssim10$, we show that $\sigma_{\ln(\bar{P})}=\ln(1+b^2\gamma^2\mathcal{M}^{4})$. This implies that thermal SZ fluctuations are easier to convert to velocity fluctuations as compared to X-ray surface brightness fluctuations. This is also true in the presence of thermal instability, as discussed in \cite{Mohapatra2019}.
	\item Density fluctuations are predominantly adiabatic in homogeneous, isotropic, unstratified turbulence. However, we find that they become increasingly isobaric in strongly stratified turbulence.
	\item The 3D velocity power spectrum is mostly independent of $\mathrm{Ri}$. However, binning along different $\mathbf{k}$ directions separately shows that $P(\mathcal{M}_{k_z})$ becomes steeper on increasing $\mathrm{Ri}$ beyond $1$. Velocity component PDFs also show significant anisotropy for $\mathrm{Ri}\gtrsim1$.
	\item  The power spectrum of density fluctuations $P(\bar{\rho}_k)$ varies both in amplitude and slope with $\mathrm{Ri}$. The slope of the power spectrum $\alpha$ initially becomes shallower with $\mathrm{Ri}$, peaks at around $\mathrm{Ri}\sim 1$ and then becomes steeper again for $\mathrm{Ri}\gtrsim 1$. This corresponds to anisotropy in turbulent eddies, becoming significant for $\mathrm{Ri}\gtrsim 1$, where $L_{\text{driv}}\approx \ell_O$ (the Ozmidov length scale). 
	\item The normalised density ($\bar{\rho}=\rho/\mean{\rho(z)}$) distribution is close to log-normal.
	
	\item We observe a positive correlation between $\delta\bar{\rho}$ and $v_z$ in stratified turbulence, for $0.01\lesssim\mathrm{Ri}\lesssim1$,  which reflects the conversion of kinetic energy into gravitational potential energy.
	
\end{itemize}

\section*{Acknowledgements}
R.~M.~acknowledges helpful discussions with Eugene Churazov, Mahendra K.~Verma and Xun Shi for sharing data for \Cref{fig:cluster-Ri-profile}. We thank the anonymous referee for helpful comments, which improved this work. R.~M.~thanks MPA Garching, SISSA Trieste, IUCAA Pune and IISc Bangalore for enabling his visits.
C.~F.~acknowledges funding provided by the Australian Research Council (Discovery Project DP170100603 and Future Fellowship FT180100495), and the Australia-Germany Joint Research Cooperation Scheme (UA-DAAD). P. S. acknowledges a Swarnajayanti Fellowship from the Department of Science and Technology, India (DST/SJF/PSA-03/2016-17), and a Humboldt fellowship for supporting his sabbatical stay at MPA Garching. We further acknowledge high-performance computing resources provided by the Leibniz Rechenzentrum and the Gauss Centre for Supercomputing (grants~pr32lo, pr48pi and GCS Large-scale project~10391), the Australian National Computational Infrastructure (grant~ek9) in the framework of the National Computational Merit Allocation Scheme and the ANU Merit Allocation Scheme. The simulation software FLASH was in part developed by the DOE-supported Flash Center for Computational Science at the University of Chicago.



\bibliographystyle{mnras}
\bibliography{refs.bib} 

\begin{thebibliography}{}
\makeatletter
\relax
\def\mn@urlcharsother{\let\do\@makeother \do\$\do\&\do\#\do\^\do\_\do\%\do\~}
\def\mn@doi{\begingroup\mn@urlcharsother \@ifnextchar [ {\mn@doi@}
  {\mn@doi@[]}}
\def\mn@doi@[#1]#2{\def\@tempa{#1}\ifx\@tempa\@empty \href
  {http://dx.doi.org/#2} {doi:#2}\else \href {http://dx.doi.org/#2} {#1}\fi
  \endgroup}
\def\mn@eprint#1#2{\mn@eprint@#1:#2::\@nil}
\def\mn@eprint@arXiv#1{\href {http://arxiv.org/abs/#1} {{\tt arXiv:#1}}}
\def\mn@eprint@dblp#1{\href {http://dblp.uni-trier.de/rec/bibtex/#1.xml}
  {dblp:#1}}
\def\mn@eprint@#1:#2:#3:#4\@nil{\def\@tempa {#1}\def\@tempb {#2}\def\@tempc
  {#3}\ifx \@tempc \@empty \let \@tempc \@tempb \let \@tempb \@tempa \fi \ifx
  \@tempb \@empty \def\@tempb {arXiv}\fi \@ifundefined
  {mn@eprint@\@tempb}{\@tempb:\@tempc}{\expandafter \expandafter \csname
  mn@eprint@\@tempb\endcsname \expandafter{\@tempc}}}

\bibitem[\protect\citeauthoryear{{Alam}, {Guha}  \& {Verma}}{{Alam}
  et~al.}{2019}]{Alam2019}
{Alam} S.,  {Guha} A.,   {Verma} M.~K.,  2019, \mn@doi [Journal of Fluid
  Mechanics] {10.1017/jfm.2019.529}, \href
  {https://ui.adsabs.harvard.edu/abs/2019JFM...875..961A} {875, 961}

\bibitem[\protect\citeauthoryear{{Aluie}}{{Aluie}}{2011}]{Aluie2011PhRvL}
{Aluie} H.,  2011, \mn@doi [\prl] {10.1103/PhysRevLett.106.174502}, \href
  {https://ui.adsabs.harvard.edu/abs/2011PhRvL.106q4502A} {106, 174502}

\bibitem[\protect\citeauthoryear{{Ar{\'e}valo}, {Churazov}, {Zhuravleva},
  {Forman}  \& {Jones}}{{Ar{\'e}valo} et~al.}{2016}]{Arevalo2016ApJ}
{Ar{\'e}valo} P.,  {Churazov} E.,  {Zhuravleva} I.,  {Forman} W.~R.,   {Jones}
  C.,  2016, \mn@doi [\apj] {10.3847/0004-637X/818/1/14}, \href
  {https://ui.adsabs.harvard.edu/abs/2016ApJ...818...14A} {818, 14}

\bibitem[\protect\citeauthoryear{{Balbus} \& {Soker}}{{Balbus} \&
  {Soker}}{1990}]{Balbus1990}
{Balbus} S.~A.,  {Soker} N.,  1990, \mn@doi [\apj] {10.1086/168926}, \href
  {https://ui.adsabs.harvard.edu/abs/1990ApJ...357..353B} {357, 353}

\bibitem[\protect\citeauthoryear{{Banerjee} \& {Galtier}}{{Banerjee} \&
  {Galtier}}{2013}]{Banerjee2013PhRvE}
{Banerjee} S.,  {Galtier} S.,  2013, \mn@doi [\pre]
  {10.1103/PhysRevE.87.013019}, \href
  {https://ui.adsabs.harvard.edu/abs/2013PhRvE..87a3019B} {87, 013019}

\bibitem[\protect\citeauthoryear{{Banerjee} \& {Galtier}}{{Banerjee} \&
  {Galtier}}{2014}]{Banerjee2014JFM}
{Banerjee} S.,  {Galtier} S.,  2014, \mn@doi [Journal of Fluid Mechanics]
  {10.1017/jfm.2013.657}, \href
  {https://ui.adsabs.harvard.edu/abs/2014JFM...742..230B} {742, 230}

\bibitem[\protect\citeauthoryear{{Banerjee} \& {Sharma}}{{Banerjee} \&
  {Sharma}}{2014}]{banerjee2014turbulence}
{Banerjee} N.,  {Sharma} P.,  2014, \mn@doi [\mnras] {10.1093/mnras/stu1179},
  \href {https://ui.adsabs.harvard.edu/\#abs/2014MNRAS.443..687B} {443, 687}

\bibitem[\protect\citeauthoryear{{Bautz} et~al.,}{{Bautz}
  et~al.}{2009}]{Bautz2009}
{Bautz} M.~W.,  et~al., 2009, \mn@doi [\pasj] {10.1093/pasj/61.5.1117}, \href
  {https://ui.adsabs.harvard.edu/abs/2009PASJ...61.1117B} {61, 1117}

\bibitem[\protect\citeauthoryear{{Boldyrev}, {Nordlund}  \&
  {Padoan}}{{Boldyrev} et~al.}{2002}]{Boldyrev2002}
{Boldyrev} S.,  {Nordlund} {\r{A}}.,   {Padoan} P.,  2002, \mn@doi [\apj]
  {10.1086/340758}, \href
  {https://ui.adsabs.harvard.edu/abs/2002ApJ...573..678B} {573, 678}

\bibitem[\protect\citeauthoryear{{Bolgiano}}{{Bolgiano}}{1962}]{Bolgiano1962}
{Bolgiano} R. J.,  1962, \mn@doi [\jgr] {10.1029/JZ067i008p03015}, \href
  {https://ui.adsabs.harvard.edu/abs/1962JGR....67.3015B} {67, 3015}

\bibitem[\protect\citeauthoryear{Bouchut, Klingenberg  \& Waagan}{Bouchut
  et~al.}{2007}]{Bouchut2007}
Bouchut F.,  Klingenberg C.,   Waagan K.,  2007, \mn@doi [Numerische
  Mathematik] {10.1007/s00211-007-0108-8}, 108, 7

\bibitem[\protect\citeauthoryear{Bouchut, Klingenberg  \& Waagan}{Bouchut
  et~al.}{2010}]{Bouchut2010}
Bouchut F.,  Klingenberg C.,   Waagan K.,  2010, \mn@doi [Numerische
  Mathematik] {10.1007/s00211-010-0289-4}, 115, 647

\bibitem[\protect\citeauthoryear{{Brethouwer} \& {Lindborg}}{{Brethouwer} \&
  {Lindborg}}{2008}]{Brethouwer2008}
{Brethouwer} G.,  {Lindborg} E.,  2008, \mn@doi [\grl] {10.1029/2007GL032906},
  \href {https://ui.adsabs.harvard.edu/abs/2008GeoRL..35.6809B} {35, L06809}

\bibitem[\protect\citeauthoryear{{Brunetti} \& {Lazarian}}{{Brunetti} \&
  {Lazarian}}{2007}]{Brunetti2007}
{Brunetti} G.,  {Lazarian} A.,  2007, \mn@doi [\mnras]
  {10.1111/j.1365-2966.2007.11771.x}, \href
  {https://ui.adsabs.harvard.edu/abs/2007MNRAS.378..245B} {378, 245}

\bibitem[\protect\citeauthoryear{{Carnevale}, {Briscolini}  \&
  {Orlandi}}{{Carnevale} et~al.}{2001}]{Carnevale2001}
{Carnevale} G.~F.,  {Briscolini} M.,   {Orlandi} P.,  2001, \mn@doi [Journal of
  Fluid Mechanics] {10.1017/S002211200000241X}, \href
  {https://ui.adsabs.harvard.edu/abs/2001JFM...427..205C} {427, 205}

\bibitem[\protect\citeauthoryear{{Cavaliere}, {Lapi}  \&
  {Fusco-Femiano}}{{Cavaliere} et~al.}{2011}]{Cavaliere2011}
{Cavaliere} A.,  {Lapi} A.,   {Fusco-Femiano} R.,  2011, \mn@doi [\aap]
  {10.1051/0004-6361/201015390}, \href
  {https://ui.adsabs.harvard.edu/abs/2011A&A...525A.110C} {525, A110}

\bibitem[\protect\citeauthoryear{{Churazov}, {Sunyaev}, {Forman}  \&
  {B{\"o}hringer}}{{Churazov} et~al.}{2002}]{Churazov2002MNRAS}
{Churazov} E.,  {Sunyaev} R.,  {Forman} W.,   {B{\"o}hringer} H.,  2002,
  \mn@doi [\mnras] {10.1046/j.1365-8711.2002.05332.x}, \href
  {https://ui.adsabs.harvard.edu/abs/2002MNRAS.332..729C} {332, 729}

\bibitem[\protect\citeauthoryear{{Churazov}, {Forman}, {Jones}  \&
  {B{\"o}hringer}}{{Churazov} et~al.}{2003}]{Churazov2003}
{Churazov} E.,  {Forman} W.,  {Jones} C.,   {B{\"o}hringer} H.,  2003, \mn@doi
  [\apj] {10.1086/374923}, \href
  {https://ui.adsabs.harvard.edu/abs/2003ApJ...590..225C} {590, 225}

\bibitem[\protect\citeauthoryear{{Churazov}, {Arevalo}, {Forman}, {Jones},
  {Schekochihin}, {Vikhlinin}  \& {Zhuravleva}}{{Churazov}
  et~al.}{2016}]{Churazov2016MNRAS}
{Churazov} E.,  {Arevalo} P.,  {Forman} W.,  {Jones} C.,  {Schekochihin} A.,
  {Vikhlinin} A.,   {Zhuravleva} I.,  2016, \mn@doi [\mnras]
  {10.1093/mnras/stw2044}, \href
  {https://ui.adsabs.harvard.edu/abs/2016MNRAS.463.1057C} {463, 1057}

\bibitem[\protect\citeauthoryear{{Corrsin}}{{Corrsin}}{1951}]{corrsin1951spectrum}
{Corrsin} S.,  1951, \mn@doi [Journal of Applied Physics] {10.1063/1.1699986},
  \href {https://ui.adsabs.harvard.edu/\#abs/1951JAP....22..469C} {22, 469}

\bibitem[\protect\citeauthoryear{{Deusebio}, {Vallgren}  \&
  {Lindborg}}{{Deusebio} et~al.}{2013}]{Deusebio2013}
{Deusebio} E.,  {Vallgren} A.,   {Lindborg} E.,  2013, \mn@doi [Journal of
  Fluid Mechanics] {10.1017/jfm.2012.611}, \href
  {https://ui.adsabs.harvard.edu/abs/2013JFM...720...66D} {720, 66}

\bibitem[\protect\citeauthoryear{{Dubey} et~al.,}{{Dubey}
  et~al.}{2008}]{Dubey2008}
{Dubey} A.,  et~al., 2008, in {Pogorelov} N.~V.,  {Audit} E.,   {Zank} G.~P.,
  eds,  Astronomical Society of the Pacific Conference Series Vol. 385,
  Numerical Modeling of Space Plasma Flows. p.~145

\bibitem[\protect\citeauthoryear{{Eswaran} \& {Pope}}{{Eswaran} \&
  {Pope}}{1988}]{eswaran1988examination}
{Eswaran} V.,  {Pope} S.~B.,  1988, Computers and Fluids, \href
  {https://ui.adsabs.harvard.edu/\#abs/1988CF.....16..257E} {16, 257}

\bibitem[\protect\citeauthoryear{{Federrath}}{{Federrath}}{2013}]{Federrath2013}
{Federrath} C.,  2013, \mn@doi [\mnras] {10.1093/mnras/stt1644}, \href
  {https://ui.adsabs.harvard.edu/\#abs/2013MNRAS.436.1245F} {436, 1245}

\bibitem[\protect\citeauthoryear{{Federrath}, {Klessen}  \&
  {Schmidt}}{{Federrath} et~al.}{2008}]{Federrath2008}
{Federrath} C.,  {Klessen} R.~S.,   {Schmidt} W.,  2008, \mn@doi [\apjl]
  {10.1086/595280}, \href
  {https://ui.adsabs.harvard.edu/abs/2008ApJ...688L..79F} {688, L79}

\bibitem[\protect\citeauthoryear{{Federrath}, {Roman-Duval}, {Klessen},
  {Schmidt}  \& {Mac Low}}{{Federrath} et~al.}{2010}]{federrath2010}
{Federrath} C.,  {Roman-Duval} J.,  {Klessen} R.~S.,  {Schmidt} W.,   {Mac Low}
  M.~M.,  2010, \mn@doi [\aap] {10.1051/0004-6361/200912437}, \href
  {https://ui.adsabs.harvard.edu/\#abs/2010A&A...512A..81F} {512, A81}

\bibitem[\protect\citeauthoryear{{Fernando} \& {Hunt}}{{Fernando} \&
  {Hunt}}{1996}]{Fernando1996}
{Fernando} H.~J.~S.,  {Hunt} J.~C.~R.,  1996, \mn@doi [Dynamics of Atmospheres
  and Oceans] {10.1016/0377-0265(95)00422-X}, \href
  {https://ui.adsabs.harvard.edu/abs/1996DyAtO..23...35F} {23, 35}

\bibitem[\protect\citeauthoryear{{Frisch}}{{Frisch}}{1995}]{Frisch1995}
{Frisch} U.,  1995, {Turbulence}

\bibitem[\protect\citeauthoryear{{Fryxell} et~al.,}{{Fryxell}
  et~al.}{2000}]{Fryxell2000}
{Fryxell} B.,  et~al., 2000, \mn@doi [The Astrophysical Journal Supplement
  Series] {10.1086/317361}, \href
  {https://ui.adsabs.harvard.edu/\#abs/2000ApJS..131..273F} {131, 273}

\bibitem[\protect\citeauthoryear{{Galtier} \& {Banerjee}}{{Galtier} \&
  {Banerjee}}{2011}]{Galtier2011PhRvL}
{Galtier} S.,  {Banerjee} S.,  2011, \mn@doi [\prl]
  {10.1103/PhysRevLett.107.134501}, \href
  {https://ui.adsabs.harvard.edu/abs/2011PhRvL.107m4501G} {107, 134501}

\bibitem[\protect\citeauthoryear{{Gaspari} \& {Churazov}}{{Gaspari} \&
  {Churazov}}{2013}]{gaspari2013constraining}
{Gaspari} M.,  {Churazov} E.,  2013, \mn@doi [\aap]
  {10.1051/0004-6361/201322295}, \href
  {https://ui.adsabs.harvard.edu/\#abs/2013A&A...559A..78G} {559, A78}

\bibitem[\protect\citeauthoryear{{Gaspari}, {Ruszkowski}  \&
  {Sharma}}{{Gaspari} et~al.}{2012}]{Gaspari2012}
{Gaspari} M.,  {Ruszkowski} M.,   {Sharma} P.,  2012, \mn@doi [\apj]
  {10.1088/0004-637X/746/1/94}, \href
  {https://ui.adsabs.harvard.edu/\#abs/2012ApJ...746...94G} {746, 94}

\bibitem[\protect\citeauthoryear{{Gaspari}, {Churazov}, {Nagai}, {Lau}  \&
  {Zhuravleva}}{{Gaspari} et~al.}{2014}]{gaspari2014}
{Gaspari} M.,  {Churazov} E.,  {Nagai} D.,  {Lau} E.~T.,   {Zhuravleva} I.,
  2014, \mn@doi [\aap] {10.1051/0004-6361/201424043}, \href
  {http://adsabs.harvard.edu/abs/2014A%26A...569A..67G} {569, A67}

\bibitem[\protect\citeauthoryear{{George}, {Fabian}, {Sanders}, {Young}  \&
  {Russell}}{{George} et~al.}{2009}]{George2009}
{George} M.~R.,  {Fabian} A.~C.,  {Sanders} J.~S.,  {Young} A.~J.,   {Russell}
  H.~R.,  2009, \mn@doi [\mnras] {10.1111/j.1365-2966.2009.14547.x}, \href
  {https://ui.adsabs.harvard.edu/abs/2009MNRAS.395..657G} {395, 657}

\bibitem[\protect\citeauthoryear{{Grete}, {O'Shea}  \& {Beckwith}}{{Grete}
  et~al.}{2020}]{Grete2020}
{Grete} P.,  {O'Shea} B.~W.,   {Beckwith} K.,  2020, \mn@doi [\apj]
  {10.3847/1538-4357/ab5aec}, \href
  {https://ui.adsabs.harvard.edu/abs/2020ApJ...889...19G} {889, 19}

\bibitem[\protect\citeauthoryear{{Herring} \& {Kimura}}{{Herring} \&
  {Kimura}}{2013}]{Herring2013}
{Herring} J.~R.,  {Kimura} Y.,  2013, \mn@doi [Physica Scripta Volume T]
  {10.1088/0031-8949/2013/T155/014031}, \href
  {https://ui.adsabs.harvard.edu/abs/2013PhST..155a4031H} {155, 014031}

\bibitem[\protect\citeauthoryear{{Hitomi Collaboration}}{{Hitomi
  Collaboration}}{2016}]{hitomi2016}
{Hitomi Collaboration} 2016, \mn@doi [\nat] {10.1038/nature18627}, \href
  {https://ui.adsabs.harvard.edu/\#abs/2016Natur.535..117H} {535, 117}

\bibitem[\protect\citeauthoryear{{Hopkins}}{{Hopkins}}{2013}]{Hopkins2013}
{Hopkins} P.~F.,  2013, \mn@doi [\mnras] {10.1093/mnras/stt010}, \href
  {https://ui.adsabs.harvard.edu/abs/2013MNRAS.430.1880H} {430, 1880}

\bibitem[\protect\citeauthoryear{{Khatri} \& {Gaspari}}{{Khatri} \&
  {Gaspari}}{2016}]{khatri2016}
{Khatri} R.,  {Gaspari} M.,  2016, \mn@doi [\mnras] {10.1093/mnras/stw2027},
  \href {http://adsabs.harvard.edu/abs/2016MNRAS.463..655K} {463, 655}

\bibitem[\protect\citeauthoryear{{Kolmogorov}}{{Kolmogorov}}{1941}]{kolmogorov1941dissipation}
{Kolmogorov} A.~N.,  1941, Akademiia Nauk SSSR Doklady, \href
  {https://ui.adsabs.harvard.edu/\#abs/1941DoSSR..32...16K} {32, 16}

\bibitem[\protect\citeauthoryear{{Konstandin}, {Girichidis}, {Federrath}  \&
  {Klessen}}{{Konstandin} et~al.}{2012}]{konstandin2012}
{Konstandin} L.,  {Girichidis} P.,  {Federrath} C.,   {Klessen} R.~S.,  2012,
  \mn@doi [\apj] {10.1088/0004-637X/761/2/149}, \href
  {https://ui.adsabs.harvard.edu/\#abs/2012ApJ...761..149K} {761, 149}

\bibitem[\protect\citeauthoryear{{Kumar}, {Chatterjee}  \& {Verma}}{{Kumar}
  et~al.}{2014}]{Kumar2014}
{Kumar} A.,  {Chatterjee} A.~G.,   {Verma} M.~K.,  2014, \mn@doi [\pre]
  {10.1103/PhysRevE.90.023016}, \href
  {https://ui.adsabs.harvard.edu/\#abs/2014PhRvE..90b3016K} {90, 023016}

\bibitem[\protect\citeauthoryear{{Lighthill}}{{Lighthill}}{1978}]{Lighthill1978}
{Lighthill} J.,  1978, {Waves in fluids}

\bibitem[\protect\citeauthoryear{{Lindborg}}{{Lindborg}}{2006}]{Lindborg2006}
{Lindborg} E.,  2006, \mn@doi [Journal of Fluid Mechanics]
  {10.1017/S0022112005008128}, \href
  {https://ui.adsabs.harvard.edu/\#abs/2006JFM...550..207L} {550, 207}

\bibitem[\protect\citeauthoryear{{Mac Low} \& {McCray}}{{Mac Low} \&
  {McCray}}{1988}]{Low1988}
{Mac Low} M.-M.,  {McCray} R.,  1988, \mn@doi [\apj] {10.1086/165936}, \href
  {https://ui.adsabs.harvard.edu/abs/1988ApJ...324..776M} {324, 776}

\bibitem[\protect\citeauthoryear{{McCourt}, {Parrish}, {Sharma}  \&
  {Quataert}}{{McCourt} et~al.}{2011}]{McCourt2011}
{McCourt} M.,  {Parrish} I.~J.,  {Sharma} P.,   {Quataert} E.,  2011, \mn@doi
  [\mnras] {10.1111/j.1365-2966.2011.18216.x}, \href
  {https://ui.adsabs.harvard.edu/abs/2011MNRAS.413.1295M} {413, 1295}

\bibitem[\protect\citeauthoryear{{Mohapatra} \& {Sharma}}{{Mohapatra} \&
  {Sharma}}{2019}]{Mohapatra2019}
{Mohapatra} R.,  {Sharma} P.,  2019, \mn@doi [\mnras] {10.1093/mnras/stz328},
  \href {https://ui.adsabs.harvard.edu/\#abs/2019MNRAS.484.4881M} {484, 4881}

\bibitem[\protect\citeauthoryear{{Mroczkowski} et~al.,}{{Mroczkowski}
  et~al.}{2019}]{Mroczkowski2019}
{Mroczkowski} T.,  et~al., 2019, \mn@doi [\ssr] {10.1007/s11214-019-0581-2},
  \href {https://ui.adsabs.harvard.edu/\#abs/2019SSRv..215...17M} {215, 17}

\bibitem[\protect\citeauthoryear{{Nelson}, {Rudd}, {Shaw}  \& {Nagai}}{{Nelson}
  et~al.}{2012}]{Nelson2012ApJ}
{Nelson} K.,  {Rudd} D.~H.,  {Shaw} L.,   {Nagai} D.,  2012, \mn@doi [\apj]
  {10.1088/0004-637X/751/2/121}, \href
  {https://ui.adsabs.harvard.edu/abs/2012ApJ...751..121N} {751, 121}

\bibitem[\protect\citeauthoryear{{Nelson}, {Lau}  \& {Nagai}}{{Nelson}
  et~al.}{2014}]{Nelson2014ApJ}
{Nelson} K.,  {Lau} E.~T.,   {Nagai} D.,  2014, \mn@doi [\apj]
  {10.1088/0004-637X/792/1/25}, \href
  {https://ui.adsabs.harvard.edu/abs/2014ApJ...792...25N} {792, 25}

\bibitem[\protect\citeauthoryear{{Nolan}, {Federrath}  \& {Sutherland}}{{Nolan}
  et~al.}{2015}]{nolan2015}
{Nolan} C.~A.,  {Federrath} C.,   {Sutherland} R.~S.,  2015, \mn@doi [\mnras]
  {10.1093/mnras/stv1030}, \href
  {http://adsabs.harvard.edu/abs/2015MNRAS.451.1380N} {451, 1380}

\bibitem[\protect\citeauthoryear{{Omma}, {Binney}, {Bryan}  \& {Slyz}}{{Omma}
  et~al.}{2004}]{Omma2004MNRAS}
{Omma} H.,  {Binney} J.,  {Bryan} G.,   {Slyz} A.,  2004, \mn@doi [\mnras]
  {10.1111/j.1365-2966.2004.07382.x}, \href
  {https://ui.adsabs.harvard.edu/abs/2004MNRAS.348.1105O} {348, 1105}

\bibitem[\protect\citeauthoryear{{Parmentier}, {Showman}  \&
  {Lian}}{{Parmentier} et~al.}{2013}]{Parmentier2013}
{Parmentier} V.,  {Showman} A.~P.,   {Lian} Y.,  2013, \mn@doi [\aap]
  {10.1051/0004-6361/201321132}, \href
  {https://ui.adsabs.harvard.edu/abs/2013A&A...558A..91P} {558, A91}

\bibitem[\protect\citeauthoryear{{Passot} \& {V{\'a}zquez-Semadeni}}{{Passot}
  \& {V{\'a}zquez-Semadeni}}{1998}]{Passot1998PhRvE}
{Passot} T.,  {V{\'a}zquez-Semadeni} E.,  1998, \mn@doi [\pre]
  {10.1103/PhysRevE.58.4501}, \href
  {https://ui.adsabs.harvard.edu/abs/1998PhRvE..58.4501P} {58, 4501}

\bibitem[\protect\citeauthoryear{Schmidt, Hillebrandt  \& Niemeyer}{Schmidt
  et~al.}{2006}]{schmidt2006numerical}
Schmidt W.,  Hillebrandt W.,   Niemeyer J.~C.,  2006, Computers \& Fluids, 35,
  353

\bibitem[\protect\citeauthoryear{{Schmidt}, {Federrath}  \&
  {Klessen}}{{Schmidt} et~al.}{2008}]{Schmidt2008}
{Schmidt} W.,  {Federrath} C.,   {Klessen} R.,  2008, \mn@doi [\prl]
  {10.1103/PhysRevLett.101.194505}, \href
  {https://ui.adsabs.harvard.edu/abs/2008PhRvL.101s4505S} {101, 194505}

\bibitem[\protect\citeauthoryear{{Schuecker}, {Finoguenov}, {Miniati},
  {B{\"o}hringer}  \& {Briel}}{{Schuecker} et~al.}{2004}]{schuecker2004}
{Schuecker} P.,  {Finoguenov} A.,  {Miniati} F.,  {B{\"o}hringer} H.,   {Briel}
  U.~G.,  2004, \mn@doi [\aap] {10.1051/0004-6361:20041039}, \href
  {http://adsabs.harvard.edu/abs/2004A%26A...426..387S} {426, 387}

\bibitem[\protect\citeauthoryear{{She} \& {Leveque}}{{She} \&
  {Leveque}}{1994}]{She1994}
{She} Z.-S.,  {Leveque} E.,  1994, \mn@doi [\prl] {10.1103/PhysRevLett.72.336},
  \href {https://ui.adsabs.harvard.edu/abs/1994PhRvL..72..336S} {72, 336}

\bibitem[\protect\citeauthoryear{{Shi} \& {Zhang}}{{Shi} \&
  {Zhang}}{2019}]{Shi2019}
{Shi} X.,  {Zhang} C.,  2019, \mn@doi [\mnras] {10.1093/mnras/stz1392}, \href
  {https://ui.adsabs.harvard.edu/abs/2019MNRAS.487.1072S} {487, 1072}

\bibitem[\protect\citeauthoryear{{Simionescu} et~al.,}{{Simionescu}
  et~al.}{2019}]{Simionescu2019SSRv}
{Simionescu} A.,  et~al., 2019, \mn@doi [\ssr] {10.1007/s11214-019-0590-1},
  \href {https://ui.adsabs.harvard.edu/abs/2019SSRv..215...24S} {215, 24}

\bibitem[\protect\citeauthoryear{{Stein}}{{Stein}}{1967}]{Stein1967}
{Stein} R.~F.,  1967, \mn@doi [\solphys] {10.1007/BF00146490}, \href
  {https://ui.adsabs.harvard.edu/abs/1967SoPh....2..385S} {2, 385}

\bibitem[\protect\citeauthoryear{{Valdarnini}}{{Valdarnini}}{2019}]{Valdarnini2019}
{Valdarnini} R.,  2019, \mn@doi [\apj] {10.3847/1538-4357/ab0964}, \href
  {https://ui.adsabs.harvard.edu/abs/2019ApJ...874...42V} {874, 42}

\bibitem[\protect\citeauthoryear{Verma}{Verma}{2018}]{verma2018}
Verma M.,  2018, Physics of Buoyant Flows.
World Scientific, New Jersey

\bibitem[\protect\citeauthoryear{{Waagan}, {Federrath}  \&
  {Klingenberg}}{{Waagan} et~al.}{2011}]{Waagan2011}
{Waagan} K.,  {Federrath} C.,   {Klingenberg} C.,  2011, \mn@doi [Journal of
  Computational Physics] {10.1016/j.jcp.2011.01.026}, \href
  {https://ui.adsabs.harvard.edu/abs/2011JCoPh.230.3331W} {230, 3331}

\bibitem[\protect\citeauthoryear{{Zeldovich} \& {Sunyaev}}{{Zeldovich} \&
  {Sunyaev}}{1969}]{Zeldovich1969}
{Zeldovich} Y.~B.,  {Sunyaev} R.~A.,  1969, \mn@doi [\apss]
  {10.1007/BF00661821}, \href
  {https://ui.adsabs.harvard.edu/abs/1969Ap&SS...4..301Z} {4, 301}

\bibitem[\protect\citeauthoryear{{Zhang}, {Churazov}  \&
  {Schekochihin}}{{Zhang} et~al.}{2018}]{Zhang2018}
{Zhang} C.,  {Churazov} E.,   {Schekochihin} A.~A.,  2018, \mn@doi [\mnras]
  {10.1093/mnras/sty1269}, \href
  {https://ui.adsabs.harvard.edu/\#abs/2018MNRAS.478.4785Z} {478, 4785}

\bibitem[\protect\citeauthoryear{{Zhuravleva}, {Churazov}, {Kravtsov}, {Lau},
  {Nagai}  \& {Sunyaev}}{{Zhuravleva} et~al.}{2013}]{zhuravleva2013quantifying}
{Zhuravleva} I.,  {Churazov} E.,  {Kravtsov} A.,  {Lau} E.~T.,  {Nagai} D.,
  {Sunyaev} R.,  2013, \mn@doi [\mnras] {10.1093/mnras/sts275}, \href
  {https://ui.adsabs.harvard.edu/\#abs/2013MNRAS.428.3274Z} {428, 3274}

\bibitem[\protect\citeauthoryear{{Zhuravleva} et~al.,}{{Zhuravleva}
  et~al.}{2014a}]{zhuravleva2014turbulent}
{Zhuravleva} I.,  et~al., 2014a, \mn@doi [\nat] {10.1038/nature13830}, \href
  {https://ui.adsabs.harvard.edu/\#abs/2014Natur.515...85Z} {515, 85}

\bibitem[\protect\citeauthoryear{{Zhuravleva} et~al.,}{{Zhuravleva}
  et~al.}{2014b}]{zhuravleva2014relation}
{Zhuravleva} I.,  et~al., 2014b, \mn@doi [\apj] {10.1088/2041-8205/788/1/L13},
  \href {https://ui.adsabs.harvard.edu/\#abs/2014ApJ...788L..13Z} {788, L13}

\bibitem[\protect\citeauthoryear{{Zhuravleva}, {Allen}, {Mantz}  \&
  {Werner}}{{Zhuravleva} et~al.}{2018}]{zhuravleva2018}
{Zhuravleva} I.,  {Allen} S.~W.,  {Mantz} A.,   {Werner} N.,  2018, \mn@doi
  [\apj] {10.3847/1538-4357/aadae3}, \href
  {https://ui.adsabs.harvard.edu/\#abs/2018ApJ...865...53Z} {865, 53}

\makeatother
\end{thebibliography}



\appendix
\renewcommand\thefigure{\thesection A\arabic{figure}}    
\section*{Appendix A}
\setcounter{figure}{0}    
\FloatBarrier
\label{sec:Appendix}
In \cref{subsec:PDFdensperturb} we showed that the density PDFs were log-normal except for a low density tail (see \Cref{fig:denshopkins}). Here we show that the density PDF is convergent over three levels of resolution so the tail is not an effect of lack of numerical resolution. 	
	
In the next plot, we justify our choice of $1024^2\times 1536$ as the resolution for most of our simulations. From both upper and lower panel of \Cref{fig:convergence-120}, we can note that the spectral slopes of Ri0.05 and Ri0.05HighRes are convergent whereas Ri0.05LowRes gives us different slopes. Doubling the resolution gives us a longer inertial range but also consumes $16\mathrm{X}$ more compute time.  This plot shows that $1024^2\times 1536$ has enough resolution to calculate velocity and density power spectra - as it gives the same slopes as the $2048^2\times 3072$.
\begin{figure}
	\includegraphics[width=\columnwidth]{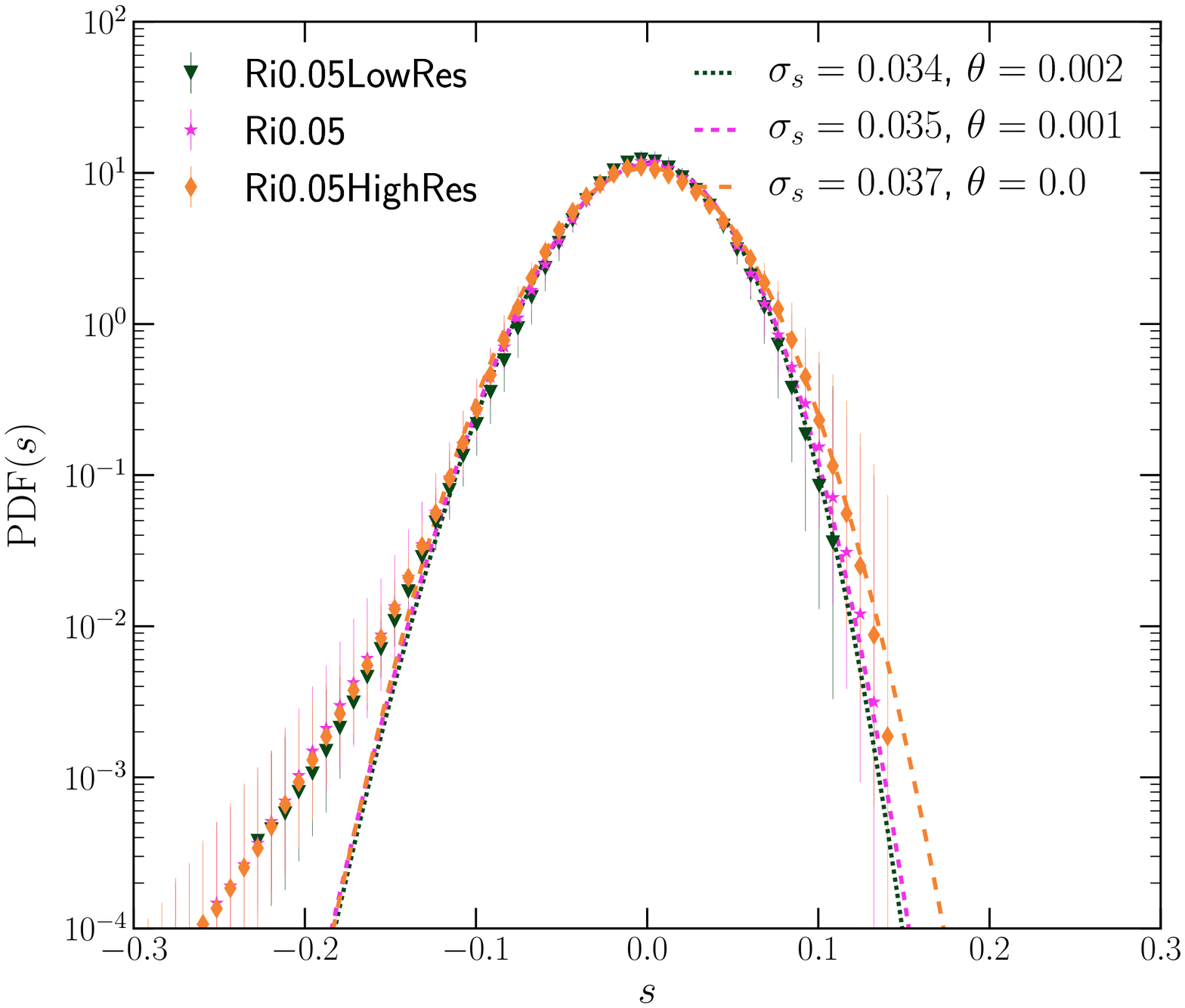}
	\hfill
	\caption [dens-PDF-convergence-check]{Volume weighted log-density PDF (similar to \Cref{fig:denshopkins}). The distributions are convergent for all three grid resolutions -- $512^2\times 768$, $1024^2\times1536$ and $2048^2\times3072$.}\label{fig:PDF-convergence-120}
\end{figure}
\begin{figure}[h]
	\includegraphics[width=\columnwidth]{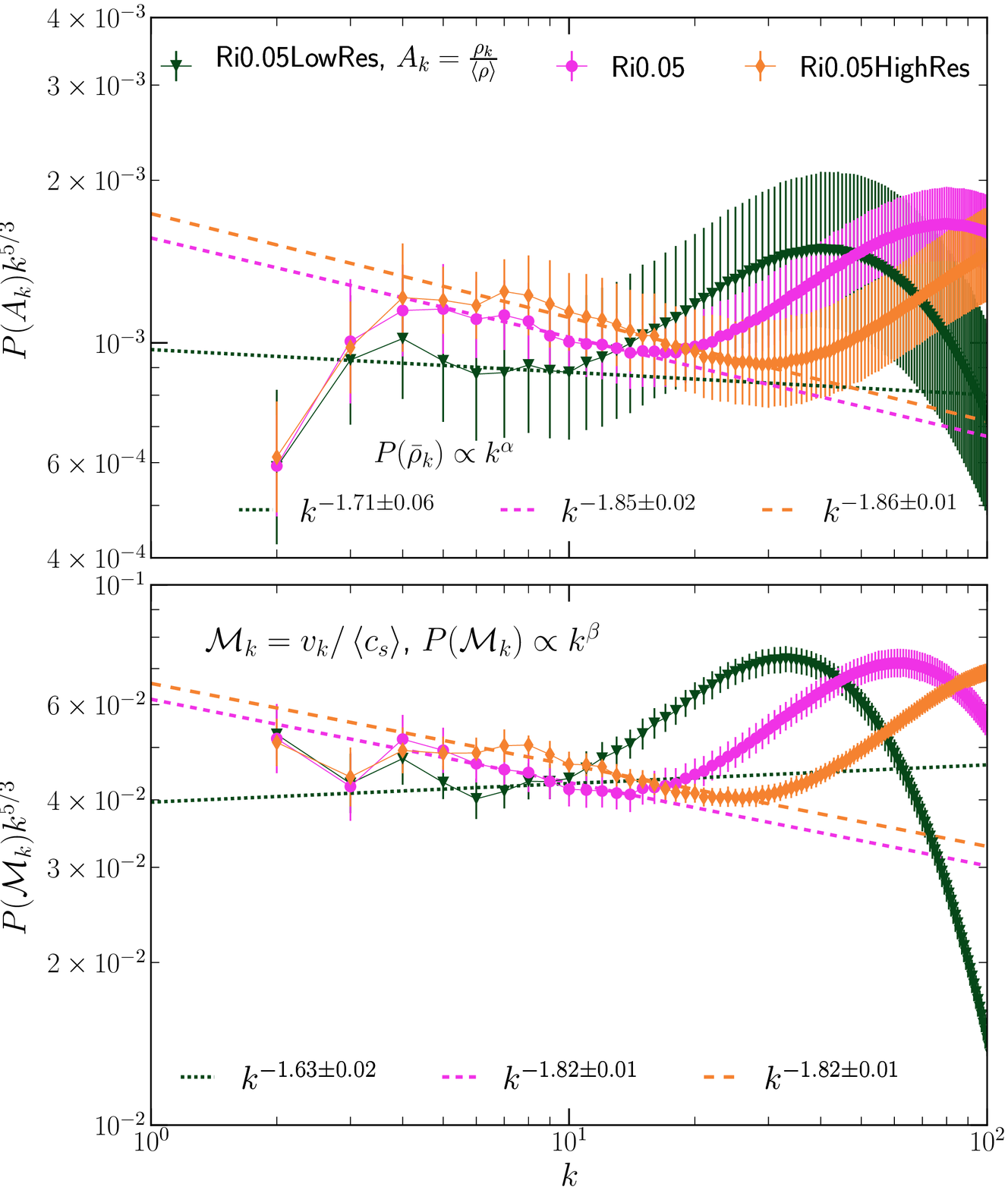}
	\hfill
	\caption [convergence-check]{Compensated power spectra of $\bar{\rho}$ (upper panel) and $\mathcal{M}$ (lower panel) for different resolutions at $\mathrm{Ri}=0.05$. The slopes converge for Ri0.05 and Ri0.05HighRes. Ri0.1LowRes has insufficient inertial range.}\label{fig:convergence-120}
\end{figure}
\section*{Additional Links}
An animation of the panels in Figure~\ref{fig:nrho-proj} is available at this URL, \href{https://www.youtube.com/watch?v=fYXbwO73Efc}{https://www.youtube.com/watch?v=fYXbwO73Efc}.

\bsp	
\label{lastpage}
\end{document}